\documentclass[11pt,a4paper]{article}
\pdfoutput=1

\usepackage{jheppub}
\usepackage{amsmath,amsfonts,amssymb,amsthm}
\usepackage{url}
\usepackage{physics}
\usepackage{hyperref}

\numberwithin{equation}{section}

\newcommand{\fft}[2]{\frac{#1}{#2}}

\newcommand{\nn}{\nonumber}

\DeclareMathAlphabet{\mathpzc}{OT1}{pzc}{m}{it}

\allowdisplaybreaks[1]

\preprint{LCTP-25-04}

\title{Multicenter higher-derivative BPS black holes}

\author[a]{Yide Cai,}
\author[a]{Sabarenath Jayaprakash,}
\author[a]{James T. Liu,}
\author[a,b]{Robert J. Saskowski}
\emailAdd{caiyi@umich.edu, sabare@umich.edu, jimliu@umich.edu, robert\_saskowski@tju.edu.cn}

\affiliation[a]{Leinweber Center for Theoretical Physics, 
University of Michigan, Ann Arbor, MI 48109, USA}
\affiliation[b]{Center for Joint Quantum Studies and Department of Physics, School of Science, Tianjin University, Tianjin 300350, China}

\abstract{We consider the reduction of four-derivative heterotic supergravity on a torus and construct two-charge multicenter BPS black hole solutions. In $d=5$, the three-form field can be dualized to a gauge field and we correspondingly construct three-charge multicenter BPS black hole solutions to the dualized Bergshoeff-de Roo action. This makes precise the embedding of known solutions into five-dimensional $\alpha'$-corrected STU supergravity.}

\date{\today}

\begin{document}

\maketitle

\section{Introduction and summary}

It is well-known that Einstein's theory of gravity is non-renormalizable and therefore not UV-complete.  Nevertheless, it serves as a natural starting point in an effective field theory expansion of the form
\begin{equation}
    \mathcal L=\mathcal L_{\partial^2}+\Lambda^{-2}\mathcal L_{\partial^4}+\mathcal O(\Lambda^{-4}),
\end{equation}
where the higher-derivative corrections encode both UV physics and quantum corrections and are suppressed by a UV scale $\Lambda$.  As string theory provides a natural UV completion to gravity, it can be studied in its low-energy limit through such an expansion.  In this case, the UV scale is naturally associated with the string length in the sense that $\alpha'=\Lambda^{-2}$, so the effective field theory is parametrized by an $\alpha'$ expansion.  As is well known, in the low-energy limit of string theory, one recovers two-derivative supergravity with higher-derivative corrections. Such corrections are also of great interest in AdS/CFT, where they are holographically dual to corrections away from the large-$N$ limit.

There has been great interest in studying black holes in string theory ever since it was realized that strings provide a natural framework for understanding quantum gravity.  Of particular interest are BPS solitons and black holes as they can be used to study non-perturbative and quantum gravity effects since they can extend between weak and strong coupling regimes of the theory.  A particular highlight of this program has been the seminal work of Strominger and Vafa~\cite{Strominger:1996sh}, which found that the Bekenstein-Hawking entropy of three-charge BPS black holes in five dimensions can be understood microscopically using a D1-D5 brane description.

While much of the study of BPS black holes has focused on the leading order two-derivative theory, higher-derivative corrections often play an important role in refining our understanding of fundamental issues.  In particular, they are crucial for exploring the weak gravity and related conjectures, for precision holography, and for developing a complete picture of black hole microstates.  Within this context, heterotic supergravity is of great interest, as its first higher-derivative correction occurs at the four-derivative level.  (This is in contrast to the Type II string where such corrections start at the eight-derivative level.)  In principle, all solutions to heterotic supergravity have been classified~\cite{Gran:2005wf,Gran:2007fu}, although this is a rather non-constructive statement. While the two-derivative solutions are often quite well understood, it is not always known what form the higher-derivative corrections take. In the past several decades, various $\alpha'$-corrected heterotic black holes have been considered (see \emph{e.g.}~\cite{Natsuume:1994hd,Kats:2006xp,Castro:2007hc,Cvitan:2007pk,Cano:2018qev,Chimento:2018kop,Cano:2018brq,Cano:2018hut,Cano:2019ycn,Ruiperez:2020qda,Cano:2021dyy,Cano:2021nzo,Ortin:2021win,Cano:2022tmn,Liu:2023fqq,Massai:2023cis}), and such solutions will be the focus of this paper. 

In many cases, rather than directly solving the dimensionally reduced effective theory, it is easier to solve the ten-dimensional equations of motion or Killing spinor equations and then interpret the solution as being reduced on a torus. At the two-derivative level, this is quite straightforward, up to a reshuffling of the fields to bring them into the standard Kaluza-Klein form. However, when higher derivatives are involved, we require nontrivial field redefinitions to identify the ten-dimensional and lower-dimensional fields, especially if we want the reduced theory to be in a canonical lower-dimensional supergravity form, such as without derivatives of field strengths.  This picture is further complicated by the freedom to perform field redefinitions on the higher-derivative couplings.  Thus, even if we are given a ten-dimensional solution, it is often still nontrivial to understand it from the lower-dimensional point of view.  Thus we will be interested in reinterpreting ten-dimensional solutions as those of an effective lower-dimensional theory.

We start by considering the dimensional reduction of the Bergshoeff-de Roo heterotic action on a $d$-dimensional torus, $T^d$.  Ignoring the original heterotic gauge fields, the resulting theory is half-maximal supergravity coupled to $d$ vector multiplets with the scalars parametrizing an $O(d,d)/O(d)\times O(d)$ coset.  The reduced four-derivative action was obtained in $O(d,d)$ form in~\cite{Eloy:2020dko,Elgood:2020xwu,Ortin:2020xdm} and in $O(d)\times O(d)$ form in~\cite{Jayaprakash:2024xlr}.  Generic black hole solutions to the reduced theory will carry $2d$ electric charges corresponding to $d$ `momentum' gauge fields and $d$ `winding' gauge fields.  However, one does not need to work with all charges turned on and can instead start from a seed solution and then use duality transformations to fill out the full charge vector.  In fact, for compactifications to five and higher dimensions, we only need to consider the fields arising from a single Kaluza-Klein circle.  As a result, we focus on the truncated theory of the lower dimensional supergravity coupled to a single vector multiplet.  Here the bosonic field content is given by a metric $g_{\mu\nu}$, an NS two-form $b_{\mu\nu}$, and a dilaton $\varphi$ along with momentum and winding gauge fields $A_\mu$ and $B_\mu$ and a Kaluza-Klein scalar $\sigma$ originating from the circle compactification. These fields are arranged into a supergravity multiplet with ($g_{\mu\nu}, b_{\mu\nu}, F_{\mu\nu}^{(-)},\varphi)$, and a vector multiplet $(F_{\mu\nu}^{(+)},\sigma)$, where $F_{\mu\nu}^{(\pm)}$ are `left' and `right' moving gauge field combinations.

In six and higher spacetime dimensions, the seed solution is a two-charge black hole.  At leading order, the BPS black hole solution can be specified by two harmonic functions $H_1$ and $H_2$, and it is striking that this harmonic function construction has a simple extension to the four-derivative case~\cite{Chimento:2018kop,Ruiperez:2020qda,Cano:2021dyy}.  For example, the gauge fields take the form (in a particular field redefinition basis)
\begin{align}
    A&=\fft1{H_1}\left(1+\fft{\alpha'}8(\partial_i\log(H_1H_2))^2\right)\dd t,\nn\\
    B&=-\fft1{H_2}\left(1+\fft{\alpha'}8(\partial_i\log(H_1H_2))^2\right)\dd t,
\label{eq:2cgauge}
\end{align}
where the partial derivatives are taken on the flat transverse space.  We demonstrate explicitly that the full solution satisfies the lower dimensional Killing spinor equations, as expected.  Indeed, we find that the Dirac structure is unchanged from the leading order and so the Killing spinor is identical to the two-derivative Killing spinor, taking into account the $\mathcal O(\alpha')$ shifted metric.

In five-dimensional spacetime, the two-form field $b$ can be dualized to a gauge field $C$ and hence the seed solution is a three-charge black hole. Such solutions were obtained previously in Ref.~\cite{Chimento:2018kop} from a ten-dimensional perspective.  Here we rewrite the solution in terms of five-dimensional supergravity fields.  In particular, while the original solution of~\cite{Chimento:2018kop} naturally corresponds to a string frame action including terms such as $(\nabla F)^2$, we use a chain of field redefinitions to obtain a final form of the BPS black hole solution corresponding to an Einstein frame supergravity action without derivatives on fields strengths.  In this case, we find the gauge fields to have the form
\begin{align}
    A = &\;\; \dfrac{1}{H_1} \Bigg(1 + \alpha' \; \dfrac{\partial_i \log\mathcal H\,\partial_i\log(H_1H_2)}{12H_3} \Bigg)\dd t, \nn \\
    B = &\;\;- \dfrac{1}{H_2} \Bigg(1 + \alpha' \; \dfrac{\partial_i \log\mathcal H\,\partial_i\log(H_1H_2)}{12H_3} \Bigg)\dd t, \nn \\
    C = &\;\;\dfrac{1}{H_3} \Bigg(1 - \alpha' \; \dfrac{\partial_i\log\mathcal H\,\partial_i\log H_3 }{12H_3}\Bigg)\dd t,
\label{eq:3cgauge}
\end{align}
where $\mathcal H=H_1H_2H_3$.  Since the torus reduction does not eliminate any supersymmetries, the black hole is a solution to $\mathcal N=4$ supergravity in five dimensions.  When $H_3=1$, this is a 1/2-BPS two-charge solution%
\footnote{The difference in the numerical factor of $\alpha'/8$ in (\ref{eq:2cgauge}) versus $\alpha'/12$ in (\ref{eq:3cgauge}) is because the former is given in string frame while the latter is in Einstein frame.}
to $\mathcal N=4$ supergravity, but when the third charge is turned on, the supersymmetry is reduced to 1/4-BPS.  If desired, one can further truncate to $\mathcal N=2$ supergravity coupled to two vector multiplets, often referred to as the STU model.  Regardless of which charges are turned on, from the perspective of $\mathcal N=2$ supergravity the black holes are 1/2-BPS solutions to the STU model.

While the full three-charge solution is presented below in (\ref{eq:finalSol}), the above form of the gauge fields is sufficient to demonstrate an important feature of the higher-derivative heterotic action.  At leading order, the three equal charge solution corresponds to taking $A=-B=C=H^{-1}\dd t$.  This allows us to truncate to a single graviphoton and hence to truncate to pure $\mathcal N=2$ supergravity without any additional vector multiplets.  However, once the higher-derivative corrections are included, this truncation is no longer possible, and the best we can do is to remove one vector multiple by setting $A=-B$.  This is an explicit demonstration that the universal vector multiplet involving the string dilaton and $b$-field cannot be consistently truncated from the full heterotic theory.

The rest of this paper is organized as follows. In Section \ref{sec:LeadingOrder}, we review the reduction of the four-derivative heterotic action and supersymmetry variations of the fermions.  We explicitly write down the truncation to a single Kaluza-Klein circle as that is all that will be needed to construct the seed solutions.  In Section \ref{sec:twoCharge}, we discuss two-charge multicenter solutions in compactified heterotic theory as well as their supersymmetry variations. In Section \ref{sec:threeCharge}, we perform the dualization of the five-dimensional action and discuss the three-charge multicenter black hole solutions, both before and after dualization. We also decompose the $\mathcal N=4$ supersymmetry variations into $\mathcal N=2$ ones and verify the Killing spinor equations. Finally, we conclude in Section \ref{sec:disc}.

%%%%%%%%%%%%%%%%%%%%%%%%%%%%%%%
\section{The reduced heterotic action}\label{sec:LeadingOrder}

The low-energy limit of heterotic string theory gives rise to a half-maximal supergravity theory in ten dimensions coupled to a set of $E_8\times E_8$ or $SO(32)$ vector multiplets.  The tree-level four-derivative corrected action has been obtained in~\cite{Metsaev:1987zx,Bergshoeff:1988nn,Bergshoeff:1989de}, and for the bosonic fields takes the form
\begin{equation}
    e^{-1}\mathcal L=e^{-2\phi}\left[R(\Omega)+4\qty(\partial_M\phi)^2-\fft1{12}\tilde H_{MNP}^2-\fft{\alpha'}8\left(\Tr F_{MN}^2-R_{MNPQ}(\Omega_+)^2\right)\right],
\label{eq:10DfourDerivAct}
\end{equation}
where
\begin{equation}
    \tilde H=\dd B+\frac{\alpha'}{4}\bigl(\omega_{3Y}(A)-\omega_{3L}(\Omega_+)\bigr),
\label{eq:Htilde}
\end{equation}
so that
\begin{equation}
    \dd\tilde H=\frac{\alpha'}{4}\bigl(\Tr F\wedge F-\Tr R(\Omega_+)\land R(\Omega_+)\bigr).
\end{equation}
Here we have introduced the torsionful connection
\begin{equation}
    \Omega_\pm=\Omega\pm\fft12\mathcal H,\qquad\mathcal H^{AB}\equiv \tilde H_M{}^{AB}\dd x^M,
\end{equation}
where $\Omega$ is the ten-dimensional torsion-free spin connection.  As highlighted in~\cite{Bergshoeff:1988nn,Bergshoeff:1989de},  this torsionful connection plays a key role in the supersymmetrization of the $\mathcal O(\alpha')$ corrections.

When the heterotic theory is reduced on a $d$-dimensional torus, $T^d$, it gives rise to half-maximal supergravity coupled to $d+16$ vector multiplets.  The string theory is invariant under the $O(d,d+16;\mathbb Z)$ T-duality group, while the supergravity scalars live on an ${O(d,d+16)/O(d)\times O(d+16)}$ coset.  The gauge fields transform as a vector of $O(d,d+16)$, while the fermions transform under $O(d)\times O(d+16)$, with the first $O(d)$ representing the R-symmetry of the theory.

For reductions to six or higher dimensions, black holes can carry electric charges under the $2d+16$ abelian gauge fields, with the charges transforming as the vector of $O(d,d+16)$.  By suitable transformations, all such configurations can be generated by starting with a two-charge solution with the two electric charges coming from the same circle in the reduction.  In five dimensions, an additional charge arises as a magnetic charge of the three-form $h$, so five-dimensional heterotic black holes carry an $O(5,21)$ vector of electric charges along with the three-form magnetic charge.  In this case, all black holes can be generated starting from a three-charge solution corresponding to a single circle reduction along with the $h$-field.  Finally, four-dimensional black holes can carry dyonic charges under the $U(1)^{28}$ gauge group and the full set of solutions can be generated starting from four-charge solutions with gauge fields on a two-torus turned on.

%4 dim BPS~\cite{Cvetic:1995bj,Cvetic:1995kv}

Here we focus on five and higher-dimensional black holes.  In this case, the most general solution can be obtained by starting with a single circle reduction and performing suitable $O(d,d+16)$ transformations on it.  However, before arriving at this system, we review some general features of the torus reduction.  The reduction of the two-derivative heterotic action on $T^d$ and verification of its $O(d,d+16;\mathbb R)$ invariance was performed in~\cite{Maharana:1992my}.  More recently, the four-derivative action was reduced in~\cite{Eloy:2020dko,Ortin:2020xdm}, and shown to be $O(d,d;\mathbb R)$ invariant in the absence of heterotic gauge fields.  Subsequently, an $O(d)\times O(d)$ invariant form of the action along with supersymmetry variations of the fermionic fields was presented in~\cite{Jayaprakash:2024xlr}.

At the two-derivative level, the standard Kaluza-Klein reduction ansatz takes the form
\begin{align}
    \dd s^2&=g_{\mu\nu}\dd x^\mu \dd x^\nu+g_{ij}\eta^i\eta^j,\qquad\eta^i=\dd y^i+A_\mu^i\dd x^\mu,\nn\\
    B&=\fft12b_{\mu\nu}\dd x^\mu\wedge \dd x^\nu+B_{\mu i}\dd x^\mu\wedge\eta^i+\fft12b_{ij}\eta^i\wedge\eta^j,\nn\\
    \phi&=\varphi+\fft14\log\det g_{ij}.
\label{eq:kkansatz}
\end{align}
Note that we have truncated the heterotic gauge fields, as they will not be needed in the construction of the two- or three-charge seed solution.  In this case, the lower-dimensional vector fields transform as an $O(d,d)$ vector
\begin{equation}
    \mathcal F=\begin{pmatrix}F^i\\G_i\end{pmatrix}=\begin{pmatrix}\dd A^i\\ \dd B_i\end{pmatrix},
\end{equation}
while the lower-dimensional scalars $g_{ij}$ and $b_{ij}$ can be organized into the $O(d,d)$ vielbein
\begin{align}
    V&=\begin{pmatrix}v^{(-)\,a}{}_i&v^{(-)\,ai}\\v^{(+)\,a}{}_i&v^{(+)\,ai}\end{pmatrix}=\begin{pmatrix}e^a{}_i+e^{aj}b_{ji}&-e^{ai}\\e^a{}_i-e^{aj}b_{ji}&e^{ai}\end{pmatrix},\nn\\
    \qquad
    V^{-1}&=\begin{pmatrix}v^{(-)\,i}{}_a&v^{(+)\,i}{}_a\\v^{(-)}_{ia}&v^{(+)}_{ia}\end{pmatrix}=\fft12\begin{pmatrix}e^i{}_a&e^i{}_a\\-e_{ia}+b_{ij}e^j{}_a&e_{ia}+b_{ij}e^j{}_a\end{pmatrix},
\end{align}
where $e_i^a$ is a vielbein for the scalar `metric' $g_{ij}$.  Note that $V^TV$ gives the $O(d,d)$ covariant scalar matrix
\begin{equation}
    V^TV=2\begin{pmatrix}g_{ij}-b_{ik}g^{kl}b_{lj}&b_{ik}g^{kj}\\-g^{ik}b_{kj}&g^{ij}\end{pmatrix}.
\end{equation}
The vielbein is a representative of the $O(d,d)/O(d)\times O(d)$ coset, and we can define the Maurer-Cartan form
\begin{equation}
    \dd VV^{-1}=\begin{pmatrix}Q^{(--)}&P^{(-+)}\\P^{(+-)}&Q^{(++)}\end{pmatrix},
\end{equation}
where the composite $O(d,d)$ connections and kinetic terms are given by
\begin{align}
    Q^{(--)\,ab}&=\fft12e^{ai}(e_i^c\dd e_j^c-e_j^c\dd e_i^c+\dd b_{ij})e^{jb},\nn\\
    Q^{(++)\,ab}&=\fft12e^{ai}(e_i^c\dd e_j^c-e_j^c\dd e_i^c-\dd b_{ij})e^{jb},\nn\\
    P^{(+-)\,ab}&=\fft12e^{ai}\dd(g_{ij}-b_{ij})e^{jb},\nn\\
    P^{(+-)\,ba}=P^{(-+)\,ab}&=\fft12e^{ai}\dd(g_{ij}+b_{ij})e^{jb}.
\end{align}

Since the $O(d,d)$ vielbein transforms between $O(d,d)$ and $O(d)\times O(d)$ indices, it can be used to define the $F^{(\pm)}$ field strengths
\begin{equation}
    \begin{pmatrix}F^{(-)}\\F^{(+)}\end{pmatrix}=V\begin{pmatrix}F^i\\G_i\end{pmatrix}=\begin{pmatrix}v^{(-)\,a}{}_iF^i+v^{(-)\,ai}G_i\\v^{(+)\,a}{}_iF^i+v^{(+)\,ai}G_i\end{pmatrix}.
\end{equation}
While the conserved electric charges transform as an $O(d,d)$ vector, supersymmetry naturally splits the field strengths into graviphotons with $F^{(-)}$ and vector multiplets with $F^{(+)}$.  Hence we will make use of the $O(d)\times O(d)$ invariant form of the $\mathcal O(\alpha')$ corrected theory given in~\cite{Jayaprakash:2024xlr}.

\subsection{The \texorpdfstring{$\mathcal O(\alpha')$}{O(alpha')} reduction}

In order to maintain $O(d)\times O(d)$ invariance as well as lower-dimensional covariance at the four-derivative level, the leading order reduction ansatz, (\ref{eq:kkansatz}), receives two corrections~\cite{Jayaprakash:2024xlr}
\begin{align}
    \delta e_i^a&=T^{ab}e_i^b,\qquad T^{ab}=\fft1{32}F_{\mu\nu}^{(-)\,a}F^{(-)\,\mu\nu\,b}+\fft14P_\mu^{(+-)\,ca}P^{(+-)\,\mu\,cb},\nn\\
    \delta B_{\mu i}&=-\frac{1}{4}\qty(\frac{1}{2}\omega_+^{\alpha\beta}F^{(-)\,c}_{\alpha\beta}+F_{\mu\alpha}^{(+)\,b}P_\alpha^{(+-)\,bc}\,\dd x^\mu)e^c_i.
\label{eq:Tab}
\end{align}
The resulting four-derivative reduced action is given in~\cite{Jayaprakash:2024xlr}, where we refer the reader for additional details.  Since we are interested in a single circle reduction, we further truncate the action of~\cite{Jayaprakash:2024xlr} by keeping only the fields along a single circle direction, which we denote as $y^9$.  In this case, only a single scalar $g_{99}=e^{2\sigma}$ survives (in addition to the dilaton) along with the momentum and winding gauge fields $A^9$ and $B_9$ with corresponding field strengths $F^9$ and $G_9$.  Furthermore, the $O(1,1)$ vielbein reduces to
\begin{equation}
    V=\begin{pmatrix}e^\sigma&-e^{-\sigma}\\e^\sigma&e^{-\sigma}\end{pmatrix},
\end{equation}
so that
\begin{equation}
    F^{(\pm)}=e^\sigma F\pm e^{-\sigma}G,\qquad P^{(+-)}=\dd\sigma.
\end{equation}
Here we have dropped the index 9 as there is no ambiguity when focusing on a single circle.

The surviving bosonic fields are the lower-dimensional metric $g_{\mu\nu}$, two scalars $\varphi$ and $\sigma$, two gauge fields $A_\mu$ and $B_\mu$, and the two-form potential $b_{\mu\nu}$.  Performing this circle truncation of the four-derivative action in~\cite{Jayaprakash:2024xlr} gives
\begin{align}
    e^{-1}\mathcal{L}= e^{-2\varphi} &\biggl[R+4(\partial_\mu\varphi)^2-(\partial_\mu\sigma)^2-\fft1{12}\tilde h_{\mu\nu\rho}^2-\fft18(F_{\mu\nu}^{(-)})^2-\fft18(F_{\mu\nu}^{(+)})^2\nn\\
    &+\fft{\alpha'}8\biggl(R_{\alpha\beta\gamma\delta}(\tilde\omega_+)^2 - R_{\alpha\beta\gamma\delta}(\tilde\omega_+) F_{\alpha\beta}^{(-)}F_{\gamma\delta}^{(-)}+ \dfrac{1}{2}  (\nabla^{(+)}_\gamma F_{\alpha\beta}^{(-)})^2 \nn\\
    &\kern3em+ \dfrac{1}{8}  F_{\alpha\beta}^{(-)}F_{\alpha\beta}^{(-)} F_{\gamma\delta}^{(-)} F_{\gamma\delta}^{(-)} 
    + \dfrac{1}{8}  F_{\alpha\beta}^{(+)} F_{\alpha\beta}^{(+)} F_{\gamma\delta}^{(+)} F_{\gamma\delta}^{(+)}\nonumber\\
    &\kern3em- \dfrac{1}{8}  F_{\alpha\beta}^{(+)}F_{\beta\gamma}^{(+)}F_{\gamma\delta}^{(+)}F_{\delta\alpha}^{(+)} +2\qty(\nabla_{[\alpha}^{'(+)}F_{\beta]\gamma}^{(+)})^2 + \qty(\partial_\gamma\sigma \, F_{\alpha\beta}^{(+)})^2\nn\\
    &\kern3em- \partial_\alpha\sigma\, \partial_\beta \sigma\, F_{\alpha\gamma}^{(+)} F_{\beta\gamma}^{(+)} + 4(\partial_\alpha\sigma\, \partial_\beta\sigma)^2 + 4\qty(\nabla_\gamma^{(+)}\partial_\alpha\sigma)^2 \nonumber\\
    & \kern3em -  R_{\alpha\beta\gamma\delta}(\omega_+) F_{\alpha\gamma}^{(+)} F_{\beta\delta}^{(+)} +\dfrac{1}{2} F_{\alpha\beta}^{(-)} F_{\beta\gamma}^{(+)} F_{\gamma\delta}^{(-)} F_{\delta\alpha}^{(+)} \nonumber\\
    &\kern3em+\dfrac{1}{4}  F_{\alpha\beta}^{(-)} F_{\beta\gamma}^{(-)} F_{\gamma\delta}^{(+)} F_{\delta\alpha}^{(+)} + \dfrac{3}{2}  (\partial_\gamma\sigma\, F_{\alpha\beta}^{(-)})^2 - 4 \partial_\gamma\sigma\, F_{\alpha\beta}^{(-)} \nabla_{[\alpha}^{'(+)} F_{\beta]\gamma}^{(+)}\nonumber\\
    &\kern3em+ 2  \partial_\beta\sigma\, F_{\alpha\gamma}^{(+)} \nabla_\gamma^{(+)} F_{\alpha\beta}^{(-)} + 2 F_{\alpha\beta}^{(-)} F_{\beta\gamma}^{(+)} \nabla_\gamma^{(+)} \partial_\alpha\sigma \biggr)\biggr],
\label{eq:4derlag}
\end{align}
where
\begin{equation}
    \tilde{h}_{\mu\nu\rho} = 3\qty(\partial_{[\mu}b_{\nu\rho]}-B_{[\mu}F_{\nu\rho]}) - \dfrac{\alpha'}{4}\left( \omega_{3L}(\tilde\omega_+)_{\mu\nu\rho}-3F_{[\mu}^{(+)\,\lambda}\nabla_{\nu\vphantom|}'^{(+)}F_{\rho]\lambda}^{(+)}\right),
\end{equation}
and the torsionful connection $\tilde\omega_\pm$ is defined recursively as
\begin{equation}
    \tilde\omega_{\pm\,\mu}{}^{\alpha\beta}=\omega_\mu{}^{\alpha\beta}\pm\fft12\tilde h_\mu{}^{\alpha\beta}.
\end{equation}
Here $\nabla_\mu^{(+)}$ is the covariant derivative with respect to the torsionful connection $\tilde\omega_+$ so that, \textit{e.g.},
\begin{equation}
    \nabla_\mu^{(+)}F_{\nu\rho}^{(-)}=\nabla_\mu^{\vphantom{(+)}} F_{\nu\rho}^{(-)}+\fft12h_{\mu\nu}^{\vphantom{(+)}}{}^\lambda F^{(-)}_{\lambda\rho}+\fft12h_{\mu\rho}^{\vphantom{(+)}}{}^\lambda F_{\nu\lambda}^{(-)},
\end{equation}
while the antisymmetrized $\nabla'^{(+)}_\mu$ covariant derivative is defined by
\begin{equation}
    \nabla_{[\mu}^{\prime(+)}F_{\nu]\rho}^{(+)}=\nabla_{[\mu}^{\vphantom{(+)}}F_{\nu]\rho}^{(+)}+\fft12h_{[\mu|\rho}^{\vphantom{(+)}}{}^\beta F_{|\nu]\beta}^{(+)}.
\end{equation}
We will obtain two- and three-charge BPS black hole solutions of this theory below.

\subsection{Supersymmetry variations}

Since we focus on BPS black holes, we will examine the Killing spinor equations corresponding to the fermionic supersymmetry variations of four-derivative action.  These supersymmetry variations were given for the torus reduction in~\cite{Jayaprakash:2024xlr} and can be written in the form
\begin{align}
    \delta_\epsilon\psi_\mu&=\Bigl[\nabla_\mu(\tilde\omega_-)-\fft14Q_\mu^{(--)\,ab}\Gamma^{ab}+\fft14F_{\mu\nu}^{(-)\,a}\qty(\delta^{ab}-\alpha'T^{ab})\Gamma^\nu\Gamma^b\nn\\
    &\quad+\fft{\alpha'}8\Bigl(\Bigl(\fft12R_{\mu\nu\alpha\beta}(\tilde\omega_+)F_{\alpha\beta}^{(-)\,a}+2P_\alpha^{(+-)\,ba}\mathcal D_{[\mu}^{\prime(+)}F_{\nu]\alpha}^{(+)\,b}-\fft14F_{\alpha\beta}^{(-)\,a}F_{\beta[\mu}^{(+)\,b}F_{\nu]\alpha}^{(+)\,b}\Bigr)\Gamma^\nu\Gamma^a\nn\\
    &\kern3em+\Bigl(P_\alpha^{(+-)\,ca}\mathcal D_\mu^{(+)}P_\alpha^{(+-)\,cb}+\fft18F_{\alpha\beta}^{(-)\,a}\mathcal D_\mu^{(+)}F_{\alpha\beta}^{(-)\,b}-\fft12F_{\alpha\beta}^{(-)\,a}F_{\mu\alpha}^{(+)\,c}P_\beta^{(+-)\,cb}\Bigr)\Gamma^{ab}\Bigr)\Bigr]\epsilon,\nn\\
    \delta_\epsilon\tilde\lambda&=\Bigl[\Gamma^\mu\partial_\mu\varphi-\fft1{12}\tilde h_{\mu\nu\rho}\Gamma^{\mu\nu\rho}+\fft18F_{\mu\nu}^{(-)\,a}\qty(\delta^{ab}-\alpha'T^{ab})\Gamma^{\mu\nu}\Gamma^b\nn\\
    &\quad+\fft{\alpha'}{16}\Bigl(\Bigl(\fft12R_{\mu\nu\alpha\beta}(\tilde\omega_+)F_{\alpha\beta}^{(-)\,a}+2P_\alpha^{(+-)\,ba}\mathcal D_{[\mu}^{\prime(+)}F_{\nu]\alpha}^{(+)\,b}-\fft14F_{\alpha\beta}^{(-)\,a}F_{\beta\mu}^{(+)\,b}F_{\nu\alpha}^{(+)\,b}\Bigr)\Gamma^{\mu\nu}\Gamma^a\nn\\
    &\kern3em+\Bigl(\fft1{12}F_{\alpha\beta}^{(-)\,a}F_{\beta\gamma}^{(-)\,b}F_{\gamma\alpha}^{(-)\,c}-F_{\alpha\beta}^{(-)\,a}P_\beta^{(+-)\,db}P_\alpha^{(+-)\,dc}\Bigr)\Gamma^{abc}\Bigr)\Bigr]\epsilon,\nn\\
    \delta_\epsilon\tilde\psi_i&=e_i^a\left[-\fft12P_\mu^{(+-)\,ab}\qty(\delta^{bc}-\alpha'T^{bc})\Gamma^\mu\Gamma^c-\fft18F_{\mu\nu}^{(+)\,a}\Gamma^{\mu\nu}\right]\epsilon.
\label{eq:fullsusy}
\end{align}
As we work in various dimensions, we have kept the spinors and Dirac matrices in their original ten-dimensional form.  In particular, the supersymmetry parameter $\epsilon$ along with the gravitino $\psi_\mu$ and gauginos $\tilde\psi_i$ are chiral ten-dimensional Majorana-Weyl spinors while the dilatino $\tilde\lambda$ has the opposite chirality.  These supersymmetry variations demonstrate that, in our conventions, $O(d)_-$ acts as the R-symmetry group while $O(d)_+$ is a flavor symmetry rotating the $d$ vector multiplets of the lower-dimensional theory.

For the case of a single circle reduction, the above variations truncate to the more compact form
\begin{align}
    \delta_\epsilon\psi_\mu&=\Bigl[\nabla_\mu(\tilde\omega_-)+\fft14F_{\mu\nu}^{(-)}(1-\alpha'T)\Gamma^\nu\Gamma^9\nn\\
    &\quad+\fft{\alpha'}8\Bigl(\fft12R_{\mu\nu\alpha\beta}(\tilde\omega_+)F_{\alpha\beta}^{(-)}+2\partial_\alpha\sigma\nabla_{[\mu}^{\prime(+)}F_{\nu]\alpha}^{(+)}-\fft14F_{\alpha\beta}^{(-)}F_{\beta[\mu}^{(+)}F_{\nu]\alpha}^{(+)}\Bigr)\Gamma^\nu\Gamma^9\Bigr]\epsilon,\nn\\
    \delta_\epsilon\tilde\lambda&=\Bigl[\Gamma^\mu\partial_\mu\varphi-\fft1{12}\tilde h_{\mu\nu\rho}\Gamma^{\mu\nu\rho}+\fft18F_{\mu\nu}^{(-)}(1-\alpha'T)\Gamma^{\mu\nu}\Gamma^9\nn\\
    &\quad+\fft{\alpha'}{16}\Bigl(\fft12R_{\mu\nu\alpha\beta}(\tilde\omega_+)F_{\alpha\beta}^{(-)}+2\partial_\alpha\sigma\nabla_{[\mu}^{\prime(+)}F_{\nu]\alpha}^{(+)}-\fft14F_{\alpha\beta}^{(-)}F_{\beta\mu}^{(+)}F_{\nu\alpha}^{(+)}\Bigr)\Gamma^{\mu\nu}\Gamma^9\Bigr]\epsilon,\nn\\
    \delta_\epsilon\tilde\psi^9&=\Bigl[-\fft12\partial_\mu\sigma(1-\alpha'T)\Gamma^\mu\Gamma^9-\fft18F_{\mu\nu}^{(+)}\Gamma^{\mu\nu}\Bigr]\epsilon,
\label{eq:susyvar}
\end{align}
where $T$ is given from (\ref{eq:Tab}) by
\begin{equation}
    T=\fft1{32}F_{\mu\nu}^{(-)}F^{(-)\,\mu\nu}+\fft14\partial_\mu\sigma\partial^\mu\sigma.
\label{eq:Ttens}
\end{equation}
%

%%%%%%%%%%%%%%%%%%%%%%%%%%%%%
\section{Two-charge BPS black holes at \texorpdfstring{$\mathcal O(\alpha')$}{O(alpha')}}\label{sec:twoCharge}

When compactified on $T^d$ to six or higher dimensions, the full set of heterotic black holes can be obtained starting from a two-charge seed solution.  However, in five dimensions, one additional charge is needed, corresponding to a magnetic three-form $h$ charge.  We first investigate the generic two-charge BPS black hole solution, and then turn to the three-charge case in the next section.

The heterotic theory compactified on $T^d$ gives rise to half-maximal supergravity coupled to $d+16$ vector multiplets.  This system admits electric black holes with charges transforming as a vector of $O(d,d+16)$.  Following the conventions of~\cite{Jayaprakash:2024xlr}, the R-symmetry is $O(d)_-$, and BPS black holes satisfy the condition $M=|Q^{(-)\,a}|$.  By a suitable duality transformation, such black holes can be brought into a two-charge form where the two charges arise from the same circle in the reduction.  Thus we can focus on the truncated lower-dimensional theory given in (\ref{eq:4derlag}) with vanishing $h$-field.

The two-charge $\alpha'$-corrected BPS black holes have been constructed in~\cite{Ruiperez:2020qda,Cano:2021dyy}, and when written in lower-dimensional variables takes the form
\begin{align}
    \dd s^2&=-\fft1{H_1H_2}(1-2\alpha'T)\,\dd t^2+\dd x^i\dd x^i,\nn\\
    \varphi&=-\fft14\log(H_1H_2)-\fft12\alpha'T,\nn\\
    \sigma&=\fft12\log(\fft{H_1}{H_2}),\nn\\
    A_t&=\fft1{H_1}(1-2\alpha'T),\nn\\
    B_t&=-\fft1{H_2}(1-2\alpha'T),
\label{eq:twocharge}
\end{align}
where 
\begin{equation}
    T=-\fft14\partial_i\log H_1\partial_i\log H_2.
\label{eq:Tcombo}
\end{equation}
In particular, this is a solution to the $\alpha'$-corrected heterotic action in the field redefinition frame given by the Lagrangian (\ref{eq:4derlag}).  This is a full multi-black hole solution in isotropic coordinates given in terms of a pair of harmonic functions, $H_1$ and $H_2$ satisfying
\begin{equation}
    \partial_i\partial_iH_1=0,\qquad\partial_i\partial_iH_2=0.
\end{equation}
Note that the higher-derivative corrections to the black hole are fully encapsulated in the correction term $T$ defined in (\ref{eq:Ttens}), as noted previously in~\cite{Cano:2018qev,Chimento:2018kop,Cano:2018brq,Ruiperez:2020qda,Cano:2021dyy}.

While the $\alpha'$-corrected two-charge solution, (\ref{eq:twocharge}), was constructed by studying the bosonic equations of motion at $\mathcal O(\alpha')$, one can also explicitly verify its supersymmetry.  After inserting the solution into the supersymmetry variations, (\ref{eq:susyvar}), we obtain
\begin{align}
    \delta_\epsilon\psi_t&=\fft12\Gamma_t{}^i\partial_if(1+\Gamma^0\Gamma^9)\epsilon,\nn\\
    \delta_\epsilon\psi_i&=e^{f/2}\left(\partial_i+\fft12\partial_if(1+\Gamma^0\Gamma^9)\right)e^{-f/2}\epsilon,\nn\\
    \delta_\epsilon\tilde\lambda&=\Gamma^i\partial_i\varphi(1+\Gamma^0\Gamma^9)\epsilon,\nn\\
    \delta_\epsilon\tilde\psi^9&=-\fft12\Gamma^i\Gamma^9\partial_i\sigma(1-\alpha'T)(1+\Gamma^0\Gamma^9)\epsilon,
\end{align}
where
\begin{equation}
    f=-\fft12\log(H_1H_2)-\alpha'T,
\end{equation}
encodes the time component of the metric, $g_{tt}=-e^{2f}$.  Since we are working in an arbitrary spacetime dimension, we have retained the ten-dimensional Dirac matrix notation of (\ref{eq:susyvar}).

Just as for the bosonic fields, (\ref{eq:twocharge}), the $\alpha'$ corrections to the supersymmetry variations are completely captured by the $T$ combination given in (\ref{eq:Tcombo}).  In particular, the Dirac structure is completely unchanged from the leading order case.  What this demonstrates is that the two-charge black hole is a 1/2-BPS configuration with Killing spinors given by
\begin{equation}
    \epsilon=e^{f/2}(1-\Gamma^0\Gamma^9)\epsilon_0,
\end{equation}
where $\epsilon_0$ is a constant spinor. Here, the $\alpha'$ dependence is hidden in the metric function $f$, which can also be seen from the fact that the (lower-dimensional) spinor bilinear $K^\mu=\bar\epsilon\Gamma^\mu\epsilon$ is a timelike Killing vector for the metric in (\ref{eq:twocharge}).

By setting $H_1=H_2=H$, the two-charge solution admits a two-equal charge truncation
\begin{align}
    \dd s^2&=-\fft1{H^2}(1-2\alpha'T)\dd t^2+\dd x^i\dd x^i,\nn\\
    \varphi&=-\fft12\log H-\fft12\alpha'T,\nn\\
    \sigma&=0,\nn\\
    A_t=-B_t&=\fft1{H}(1-2\alpha'T),
\end{align}
where
\begin{equation}
    T=-\fft14(\partial_i\log H)^2.
\label{eq:2eqT}
\end{equation}
This is a solution to the pure half-maximal supergravity that one gets after truncating away the vector multiplets~\cite{Liu:2023fqq}.

\subsection{Field redefinitions and other forms of the solution}

It should be noted that, when working perturbatively in the higher-derivative expansion, one may shift expressions around by performing field redefinitions.  Thus the form of the Lagrangian, as well as solutions, is hardly unique.  Depending on the situation, some field redefinition frames may be more convenient than others.  Nevertheless, they are all physically equivalent, although care must be taken when comparing results.

The BPS black hole solution, (\ref{eq:twocharge}), is a solution to the $\mathcal O(\alpha')$ corrected Lagrangian, (\ref{eq:4derlag}), that was directly obtained by compactification of the ten-derivative heterotic Lagrangian given in the form of (\ref{eq:10DfourDerivAct}).   In particular, the compactification of the ten-dimensional Riemann-squared term gives rise to the $(\nabla F)^2$ term in (\ref{eq:4derlag}).  In many cases, it is more natural to work in a field redefinition frame that removes such derivatives of field strengths.  For the toroidal reduction, this was performed in~\cite{Jayaprakash:2024xlr}, and the resulting Lagrangian specialized to a circle reduction takes the form
\begin{align}
    e^{-1}\mathcal{L}= e^{-2\varphi} &\biggl[R+4(\partial_\mu\varphi)^2-(\partial_\mu\sigma)^2-\fft1{12}\tilde h_{\mu\nu\rho}^2-\fft18(F_{\mu\nu}^{(-)})^2-\fft18(F_{\mu\nu}^{(+)})^2\nn\\
    &+\fft{\alpha'}8\biggl(R_{\alpha\beta\gamma\delta}(\tilde\omega_+)^2 - \fft12R^{\alpha\beta\gamma\delta}(\tilde\omega_+) F_{\alpha\beta}^{(-)}F_{\gamma\delta}^{(-)} - \dfrac{1}{8}  F_{\alpha\beta}^{(+)} F_{\alpha\beta}^{(+)} F_{\gamma\delta}^{(+)} F_{\gamma\delta}^{(+)}\nonumber\\
    &\kern3em+ \dfrac{1}{8}  F_{\alpha\beta}^{(+)}F_{\beta\gamma}^{(+)}F_{\gamma\delta}^{(+)}F_{\delta\alpha}^{(+)}  -\fft12  (F_{\alpha\beta}^{(+)})^2(\partial_\gamma\sigma)^2+  F_{\alpha\gamma}^{(+)} F_{\beta\gamma}^{(+)}\,\partial^\alpha\sigma\, \partial^\beta \sigma\nonumber\\
    & \kern3em  +\dfrac{1}{8} F_{\alpha\beta}^{(+)} F_{\alpha\beta}^{(-)} F_{\gamma\delta}^{(+)} F_{\gamma\delta}^{(-)}  -\dfrac{1}{4} F_{\alpha\beta}^{(-)} F_{\beta\gamma}^{(+)} F_{\gamma\delta}^{(-)} F_{\delta\alpha}^{(+)} \nonumber\\
    &\kern3em-\dfrac{1}{4}  F_{\alpha\beta}^{(-)} F_{\beta\gamma}^{(-)} F_{\gamma\delta}^{(+)} F_{\delta\alpha}^{(+)} + \dfrac{1}{2}  (F_{\alpha\beta}^{(-)})^2(\partial_\gamma\sigma)^2 -  F_{\alpha\gamma}^{(-)} F_{\beta\gamma}^{(-)}\,\partial^\alpha\sigma\, \partial^\beta \sigma\nonumber\\
    &\kern3em+\fft12h^{\alpha\beta\gamma}F^{(+)}_{\alpha\delta}F^{(-)}_{\beta\gamma}\partial_\delta\sigma+\fft12h^{\alpha\beta\gamma}F^{(-)}_{\alpha\delta}F^{(+)}_{\beta\gamma}\partial_\delta\sigma-h^{\alpha\beta\gamma}F^{(+)}_{\alpha\delta}F^{(-)}_{\gamma\delta}\partial_\beta\sigma
    \biggr)\biggr],
\label{eq:4derlag2}
\end{align}
where now
\begin{equation}
    \tilde{h}_{\mu\nu\rho} = 3(\partial_{[\mu}b_{\nu\rho]}-B_{[\mu}F_{\nu\rho]}) - \dfrac{\alpha'}{4} \omega_{3L}(\tilde\omega_+)_{\mu\nu\rho}.
\end{equation}

The Lagrangian, (\ref{eq:4derlag2}), can be obtained from (\ref{eq:4derlag}) by the field redefinitions of the form $\Phi\to\Phi+\frac{\alpha'}{8}\delta\Phi$ where
\begin{align}
    \delta g_{\mu\nu}&=F_{\mu\lambda}^{(-)\,a}F_{\nu\lambda}^{(-)\,a}+F_{\mu\lambda}^{(+)\,a}F_{\nu\lambda}^{(+)\,a}+4P_\mu^{(+-)\,ab}P_\nu^{(+-)\,ab},\nn\\
    \delta\varphi&=\fft14F_{\mu\nu}^{(-)\,a}F_{\mu\nu}^{(-)\,a}+\fft14F_{\mu\nu}^{(+)\,a}F_{\mu\nu}^{(+)\,a}+P_\mu^{(+-)\,ab}P_\mu^{(+-)\,ab},\nn\\
    \delta A_\mu^i&=\dfrac{1}{2}v^{(-)\,i}{}_a\qty(4F_{\mu\nu}^{(+)\,b}P_\nu^{(+-)\,ba}+h_{\mu}{}^{\nu\lambda}F_{\nu\lambda}^{(-)\,a}),\nn\\
    \delta B_{\mu i}&=\dfrac{1}{2}v^{(-)}_{ia}\qty(4F_{\mu\nu}^{(+)\,b}P_\nu^{(+-)\,ba}+h_{\mu}{}^{\nu\lambda}F_{\nu\lambda}^{(-)\,a}),\nn\\
    \delta b_{\mu\nu}&=B_{[\nu|i}\delta A_{|\mu]}^i.
\end{align}
For the circle reduction, this reduces to
\begin{align}
    \delta g_{\mu\nu}&=F_{\mu\lambda}^{(-)}F_{\nu\lambda}^{(-)}+F_{\mu\lambda}^{(+)}F_{\nu\lambda}^{(+)}+4\partial_\mu\sigma\partial_\nu\sigma,\nn\\
    \delta\varphi&=\fft14F_{\mu\nu}^{(-)}F_{\mu\nu}^{(-)}+\fft14F_{\mu\nu}^{(+)}F_{\mu\nu}^{(+)}+(\partial_\mu\sigma)^2,\nn\\
    \delta A_\mu&=\dfrac{1}{2}e^{-\sigma}\qty(4F_{\mu\nu}^{(+)}\partial^\nu\sigma+h_{\mu}{}^{\nu\lambda}F_{\nu\lambda}^{(-)}),\nn\\
    \delta B_\mu&=-\dfrac{1}{2}e^\sigma\qty(4F_{\mu\nu}^{(+)}\partial^\nu\sigma+h_{\mu}{}^{\nu\lambda}F_{\nu\lambda}^{(-)}),\nn\\
    \delta b_{\mu\nu}&=B_{[\nu}\delta A_{\mu]}.
\end{align}
Performing these field redefinitions on the two-charge solution, (\ref{eq:twocharge}) gives the field-redefined BPS black hole
\begin{align}
    \dd s^2&=-\fft1{H_1H_2}\qty(1-\fft{\alpha'}4\qty(\partial_i\log(H_1/H_2))^2)\dd t^2+\dd x^i\dd x^i-\fft{\alpha'}8\qty(\partial_i\log(H_1H_2)\dd x^i)^2,\nn\\
    \varphi&=-\fft14\log(H_1H_2)-\fft{\alpha'}{32}\qty(\qty(\partial_i\log(H_1H_2))^2+2\qty(\partial_i\log(H_1/H_2))^2),\nn\\
    \sigma&=\fft12\log(\fft{H_1}{H_2}),\nn\\
    A_t&=\fft1{H_1}\qty(1+\fft{\alpha'}8(\partial_i\log(H_1H_2))^2),\nn\\
    B_t&=-\fft1{H_2}\qty(1+\fft{\alpha'}8(\partial_i\log(H_1H_2))^2).
\label{eq:fr2cbh}
\end{align}
This is a solution to the Lagrangian (\ref{eq:4derlag2}) written without derivatives of field strengths.  Note that, after field redefinitions, this metric no longer has the form of an isotropic coordinate solution.

Note that the $\alpha'$ corrections in the field-redefined solution (\ref{eq:fr2cbh}) are organized in terms of logs of the product and ratio of the leading order harmonic functions.  The product $(H_1H_2)$ represents the graviphoton charge while the ratio $(H_1/H_2)$ represents the vector multiplet charge.  Setting $H_1=H_2=H$ gives the two-equal charge solution
\begin{align}
    \dd s^2&=-\fft1{H^2}\dd t^2+\dd x^i\dd x^i-\fft{\alpha'}2\left(\partial_i\log H\, \dd x^i\right)^2,\nn\\
    \varphi&=-\fft12\log H+\fft{\alpha'}2T,\nn\\
    \sigma&=0,\nn\\
    A_t&=\fft1{H}(1-2\alpha'T),\nn\\
    B_t&=-\fft1{H}(1-2\alpha'T),
\end{align}
where $T$ is unchanged from (\ref{eq:2eqT}).

There is one more field redefinition frame to note: the one with a torsion-free spin connection.  The torsion-free Lagrangian can be obtained from the torsionful one, (\ref{eq:4derlag2}), by a field redefinition
\begin{align}
    \delta g_{\mu\nu} &= h_{\mu\rho\sigma} h_{\nu}^{\ \ \rho\sigma},\nn\\
    \delta \varphi \ \ &= \dfrac{1}{4}  h_{\mu\nu\sigma} h^{\mu\nu\sigma}. \label{eq:fredef2}
\end{align}
The resulting Lagrangian takes the form
\begin{align}
    e^{-1}\mathcal{L}= e^{-2\varphi} &\biggl[R+4(\partial_\mu\varphi)^2-(\partial_\mu\sigma)^2-\fft1{12}\tilde h_{\mu\nu\rho}^2-\fft18(F_{\mu\nu}^{(-)})^2-\fft18(F_{\mu\nu}^{(+)})^2\nn\\
    &+\fft{\alpha'}8\biggl(R_{\alpha\beta\gamma\delta}^2 - \fft12R^{\alpha\beta\gamma\delta}F_{\alpha\beta}^{(-)}F_{\gamma\delta}^{(-)}-\fft18(h_{\mu\nu}^2)^2+\fft1{24}h^4-\fft12R_{\alpha\beta\gamma\delta}h^{\alpha\beta\rho}h^{\gamma\delta}{}_\rho\nn\\
    &\kern3em +\dfrac{1}{32}  F_{\alpha\beta}^{(-)} F_{\alpha\beta}^{(-)} F_{\gamma\delta}^{(-)} F_{\gamma\delta}^{(-)}-\fft1{16}F_{\alpha\beta}^{(-)}F_{\beta\gamma}^{(-)}F_{\gamma\delta}^{(-)}F_{\delta\alpha}^{(-)}+\fft1{16}h^{\alpha\beta\rho}h^{\gamma\delta}{}_\rho F_{\alpha\beta}^{(-)}F_{\gamma\delta}^{(-)}\nn\\
    &\kern3em+\fft18h^{\alpha\beta\rho}h^{\gamma\delta}{}_\rho F_{\alpha\gamma}^{(-)}F_{\beta\delta}^{(-)}-\fft14h^2_{\mu\nu}F_{\mu\beta}^{(-)}F_{\nu\beta}^{(-)}- \dfrac{1}{32}  F_{\alpha\beta}^{(+)} F_{\alpha\beta}^{(+)} F_{\gamma\delta}^{(+)} F_{\gamma\delta}^{(+)}\nn\\
    &\kern3em- \dfrac{1}{16}  F_{\alpha\beta}^{(+)}F_{\beta\gamma}^{(+)}F_{\gamma\delta}^{(+)}F_{\delta\alpha}^{(+)}-\fft12  (F_{\alpha\beta}^{(+)})^2(\partial_\gamma\sigma)^2+  F_{\alpha\gamma}^{(+)} F_{\beta\gamma}^{(+)}\partial^\alpha\sigma \partial^\beta \sigma\nonumber\\
    & \kern3em  -\dfrac{1}{4} F_{\alpha\beta}^{(-)} F_{\beta\gamma}^{(-)} F_{\gamma\delta}^{(+)} F_{\delta\alpha}^{(+)} -h^2_{\mu\nu}\partial^\mu\sigma\partial^\nu\sigma-\fft14h^2_{\mu\nu}F_{\mu\beta}^{(+)}F_{\nu\beta}^{(+)}\nonumber\\
    &\kern3em-\fft1{16}h^{\alpha\beta\rho}h^{\gamma\delta}{}_\rho F_{\alpha\beta}^{(+)}F_{\gamma\delta}^{(+)}+\fft18h^{\alpha\beta\rho}h^{\gamma\delta}{}_\rho F_{\alpha\gamma}^{(+)}F_{\beta\delta}^{(+)}+\fft12(F_{\alpha\beta}^{(-)})^2(\partial_\gamma\sigma)^2\nn\\
    &\kern3em- F_{\alpha\gamma}^{(-)} F_{\beta\gamma}^{(-)}\partial^\alpha\sigma \partial^\beta \sigma+\fft12h^{\alpha\beta\gamma}F^{(+)}_{\alpha\delta}F^{(-)}_{\beta\gamma}\partial^\delta\sigma+\fft12h^{\alpha\beta\gamma}F^{(-)}_{\alpha\delta}F^{(+)}_{\beta\gamma}\partial^\delta\sigma\nn\\
    &\kern3em-h^{\alpha\beta\gamma}F^{(+)}_{\alpha\delta}F^{(-)}_{\gamma\delta}\partial_\beta\sigma
    \biggr)\biggr],
\label{eq:4derlag3}
\end{align}
where
\begin{equation}
    \tilde{h}_{\mu\nu\rho} = 3(\partial_{[\mu}b_{\nu\rho]}-B_{[\mu}F_{\nu\rho]}) - \dfrac{\alpha'}{4} \omega_{3L}(\omega)_{\mu\nu\rho},
\end{equation}
and we have defined $h^2_{\mu\nu}=h_{\mu\rho\lambda}h_\nu{}^{\rho\lambda}$. This Lagrangian, written in terms of a torsion-free connection, can be directly compared with that of~\cite{Eloy:2020dko}.

Since the two-charge solution, (\ref{eq:fr2cbh}), does not make use of the three-form $h$, it is unaffected by the field redefinitions (\ref{eq:fredef2}), and hence remains a solution to the torsion-free Lagrangian, (\ref{eq:4derlag3}).  However, this will no longer be the case for string solutions or three-charge black holes in five dimensions that involve the $h$-field.

%%%%%%%%%%%%%%%%%%%%%%%%%%%%%%%%%%%%%%%%%%
\section{Three-charge BPS black holes in five dimensions}\label{sec:threeCharge}

In five dimensions, the three-form $h$ can support a magnetically charged black hole.  Equivalently, it can be dualized to a two-form, thus allowing for an additional electric charge.  By appropriate duality transformations, all black hole solutions to heterotic theory on $T^5$ can be obtained starting from a three-charge solution with the three charges corresponding to momentum and winding charges for a circle reduction along with a magnetic three-form charge.

The Lagrangian (\ref{eq:4derlag}) admits a three-charge BPS black hole solution of the form~\cite{Chimento:2018kop}
\begin{align}
    \dd s^2&=-\fft1{H_1H_2}(1-2\alpha'T)\dd t^2+H_3\qty(1+\alpha'\fft{(\partial_i\log H_3)^2}{4H_3})\dd x^i\dd x^i,\nn\\
    \varphi&=\fft14\log(\fft{(H_3)^2}{H_1H_2})-\fft12\alpha'\left(T-\fft{(\partial_i\log H_3)^2}{4H_3}\right),\nn\\
    \sigma&=\fft12\log(\fft{H_1}{H_2}),\nn\\
    A_t&=\fft1{H_1}(1-2\alpha'T),\nn\\
    B_t&=-\fft1{H_2}(1-2\alpha'T),\nn\\
    h_{ijk}&=\epsilon_{ijkl}\partial_lH_3,
\label{eq:3chargesol}
\end{align}
where $\partial_i\partial_iH_a=0$ and
\begin{equation}
    T=-\fft1{4H_3}\partial_i\log H_1\partial_i\log H_2.
\end{equation}
Compared to the two-charge solution, (\ref{eq:twocharge}), the additional charge is encoded in the third harmonic function, $H_3$.  The inverse $H_3$ factor in the definition of $T$ is just the inverse metric $g^{ij}=\delta^{ij}/H_3$ that contracts the two spatial derivatives.  There is also an explicit $\mathcal O(\alpha')$ correction to the metric and dilaton proportional to $(\partial_i\log H_3)^2$ that vanishes when the third charge is removed.  Note, however, that the three-form field strength $h_{ijk}$ does not receive any $\mathcal O(\alpha')$ corrections.

We can also verify that the three-charge solutions remain supersymmetric at $\mathcal O(\alpha')$.  Insertion of the solution into the supersymmetry variations, (\ref{eq:susyvar}), gives
\begin{align}
    \delta_\epsilon\psi_t&=\fft12\Gamma_t{}^i\partial_if(1+\Gamma^0\Gamma^9)\epsilon,\nn\\
    \delta_\epsilon\psi_i&=e^{f/2}\biggl[\partial_i+\fft12\partial_if(1+\Gamma^0\Gamma^9)+\fft12\Gamma_i{}^j\partial_jg((1+\Gamma^0\Gamma^9)+\Gamma^{09}(1-\Gamma^{5678}\Gamma^{11}))\nn\\
    &\kern15.3em-\fft{\alpha'}8\Gamma_i{}^j\fft1{H_3}\partial_j\log H_3\fft{\partial_k\partial_kH_3}{H_3}\Gamma^{09}\Gamma^{5678}\Gamma^{11}\biggr]e^{-f/2}\epsilon,\nn\\
    \delta_\epsilon\tilde\lambda&=\Gamma^i\biggl[\partial_i\varphi(1+\Gamma^0\Gamma^9)-\partial_ig\,\Gamma^{09}(1-\Gamma^{5678}\Gamma^{11})-\fft{\alpha'}4\fft1{H_3}\partial_i\log H_3\fft{\partial_j\partial_jH_3}{H_3}\Gamma^{09}\Gamma^{5678}\Gamma^{11}\biggr]\epsilon,\nn\\
    \delta_\epsilon\tilde\psi^9&=-\fft12\Gamma^i\Gamma^9\partial_i\sigma(1-\alpha'T)(1+\Gamma^0\Gamma^9)\epsilon,
\label{eq:3Qsusy}
\end{align}
where $\Gamma^{11}=\Gamma^{012\cdots9}$.  As in the above, since we started with an arbitrary $d$-dimensional torus reduction, we have retained a ten-dimensional notation for the spinors and Dirac matrices.  However, when focusing on the five-dimensional reduction, it makes sense to shift to a five-dimensional spinor notation.  We will consider this below.  But first, a few comments are in order.

To interpret the supersymmetry variations, (\ref{eq:3Qsusy}), we note that the function $H_3$ is harmonic so that in particular $\partial_i\partial_iH_3=0$.  Unlike for $H_1$ and $H_2$, where the harmonic function condition comes from the corresponding equations of motion, the $H_3$ harmonic function condition shows up directly in the Killing spinor equations.  The distinction arises because $H_3$ corresponds to a magnetic three-form charge where harmonicity comes from the Bianchi identity and not the equation of motion.

Assuming $H_3$ is harmonic, the remaining structure of the supersymmetry variations involves the two commuting projections
\begin{equation}
    P_1=\fft12(1+\Gamma^0\Gamma^9),\qquad P_2=\fft12(1-\Gamma^{5678}\Gamma^{11}).
\end{equation}
The Killing spinor then takes the form
\begin{equation}
    \epsilon=e^{f/2}(1-\Gamma^0\Gamma^9)(1+\Gamma^{5678}\Gamma^{11})\epsilon_0,
\end{equation}
and the three-charge black hole preserves $1/4$ of the original heterotic supersymmetries, corresponding to four real supercharges out of 16.  Equivalently, from a five-dimensional perspective, this black hole is a $1/4$-BPS solution to half-maximal $\mathcal N=4$ supergravity.  The supersymmetry is enhanced to $1/2$-BPS in the two-charge limit when the third $H_3$ charge is turned off.

\subsection{Five-dimensional \texorpdfstring{$\mathcal N=4$}{N=4} supergravity and truncation to \texorpdfstring{$\mathcal N=2$}{N=2}}

Since here we focus on five dimensions, it is instructive to reduce the ten-dimensional spinors to five dimensions.  We thus introduce the decomposition
\begin{align}
    \Gamma^{\alpha}&=\gamma^\mu\otimes\mathbf1\otimes\sigma^1\kern4em(\mu=0,\ldots,4),\nn\\
    \Gamma^{a}&=\mathbf1\otimes t^{a-4}\otimes\sigma^2\kern3.3em(a=5,\ldots,9).
\end{align}
In addition, we use the conventions that
\begin{equation}
    \gamma_{01234}=i,\qquad t^{12345}=1,\qquad\Gamma^{11}\equiv\Gamma^{0\cdots9}=\mathbf1\otimes\mathbf1\otimes\sigma^3.
\end{equation}
With this definition of $\Gamma^{11}$, the Majorana-Weyl condition in ten dimensions indicates that spinors have a definite chirality under $\sigma^3$.  We thus let (with a slight abuse of notation)
\begin{equation}
    \epsilon\to\begin{pmatrix}\epsilon\\0\end{pmatrix},\qquad\psi_\mu\to\begin{pmatrix}\psi_\mu\\0\end{pmatrix},\qquad\tilde\psi^a\to\begin{pmatrix}\chi_a\\0\end{pmatrix},\qquad\tilde\lambda\to\begin{pmatrix}0\\i\lambda\end{pmatrix}.
\end{equation}
While these spinors can be further reduced to a set of four symplectic-Majorana fermions in five dimensions, the specifics of this will not be needed for our discussion.  We note, however, that the `internal' Dirac matrices $t^a$ are used to relate the spinor and vector representations of the $SO(5)_-\simeq Sp(4)$ R-symmetry group of the five-dimensional $\mathcal N=4$ theory.

The full fermionic supersymmetry variations, (\ref{eq:fullsusy}), then reduce to
\begin{align}
    \delta_\epsilon\psi_\mu&=\Bigl[\nabla_\mu(\tilde\omega_-)-\fft14Q_\mu^{(--)\,ab}t^{ab}+\fft{i}4F_{\mu\nu}^{(-)\,a}\qty(\delta^{ab}-\alpha'T^{ab})\gamma^\nu t^b\nn\\
    &\quad+\fft{\alpha'}8\Bigl(i\Bigl(\fft12R_{\mu\nu\alpha\beta}(\tilde\omega_+)F_{\alpha\beta}^{(-)\,a}+2P_\alpha^{(+-)\,ba}\mathcal D_{[\mu}^{\prime(+)}F_{\nu]\alpha}^{(+)\,b}-\fft14F_{\alpha\beta}^{(-)\,a}F_{\beta[\mu}^{(+)\,b}F_{\nu]\alpha}^{(+)\,b}\Bigr)\gamma^\nu t^a\nn\\
    &\kern3em+\Bigl(P_\alpha^{(+-)\,ca}\mathcal D_\mu^{(+)}P_\alpha^{(+-)\,cb}+\fft18F_{\alpha\beta}^{(-)\,a}\mathcal D_\mu^{(+)}F_{\alpha\beta}^{(-)\,b}-\fft12F_{\alpha\beta}^{(-)\,a}F_{\mu\alpha}^{(+)\,c}P_\beta^{(+-)\,cb}\Bigr)t^{ab}\Bigr)\Bigr]\epsilon,\nn\\
    \delta_\epsilon\lambda&=\Bigl[-i\gamma^\mu\partial_\mu\varphi+\fft{i}{12}\tilde h_{\mu\nu\rho}\gamma^{\mu\nu\rho}+\fft18F_{\mu\nu}^{(-)\,a}\qty(\delta^{ab}-\alpha'T^{ab})\gamma^{\mu\nu}t^b\nn\\
    &\quad+\fft{\alpha'}{16}\Bigl(\Bigl(\fft12R_{\mu\nu\alpha\beta}(\tilde\omega_+)F_{\alpha\beta}^{(-)\,a}+2P_\alpha^{(+-)\,ba}\mathcal D_{[\mu}^{\prime(+)}F_{\nu]\alpha}^{(+)\,b}-\fft14F_{\alpha\beta}^{(-)\,a}F_{\beta\mu}^{(+)\,b}F_{\nu\alpha}^{(+)\,b}\Bigr)\gamma^{\mu\nu}t^a\nn\\
    &\kern3em+\Bigl(\fft1{12}F_{\alpha\beta}^{(-)\,a}F_{\beta\gamma}^{(-)\,b}F_{\gamma\alpha}^{(-)\,c}-F_{\alpha\beta}^{(-)\,a}P_\beta^{(+-)\,db}P_\alpha^{(+-)\,dc}\Bigr)t^{abc}\Bigr)\Bigr]\epsilon,\nn\\
    \delta_\epsilon\chi^a&=\left[-\fft{i}2\gamma^\mu P_\mu^{(+-)\,ab}\qty(\delta^{bc}-\alpha'T^{bc})t^c-\fft18F_{\mu\nu}^{(+)\,a}\gamma^{\mu\nu}\right]\epsilon.
\label{eq:redsusy}
\end{align}
The five-dimensional theory is that of $\mathcal N=4$ supergravity with gravity multiplet fields $(g_{\mu\nu},\psi_\mu,F_{\mu\nu}^{(-)\,a},h_{\mu\nu\lambda},\lambda,\varphi)$ coupled to five vector multiplets with fields $(F_{\mu\nu}^{(+)\,b},\chi^b,P_\mu^{(+-)\,ba})$.  Here $a$ is a vector index of the $Sp(4)$ R-symmetry group while $b$ is a flavor index transforming as a vector of $SO(5)_+$.

The $\mathcal N=4$ fields can be decomposed in terms of $\mathcal N=2$ multiplets.  Corresponding to our choice of only turning on charges in the $y^9$ direction, we focus on an $\mathcal N=2$ subset of the full $\mathcal N=4$ supersymmetries by choosing the supersymmetry parameter to satisfy
\begin{equation}
    \epsilon=t^5\epsilon.
\end{equation}
In this case, the $\mathcal N=4$ supergravity multiplet decomposes into an $\mathcal N=2$ supergravity multiplet with fields $(g_{\mu\nu},\tilde\psi_\mu^{(+)},F_{\mu\nu}^{(-)\,9}-H_{\mu\nu})$, along with an $\mathcal N=2$ gravitino multiplet with fields $(\psi_\mu^{(-)},F_{\mu\nu}^{(-)\,\hat a},\lambda^{(-)})$ and an $\mathcal N=2$ vector multiplet with fields $(F_{\mu\nu}^{(-)\,9}+2H_{\mu\nu},\lambda^{(+)},\varphi)$.  Here the two-form $H$ is the Hodge dual of the three-form $h$.  Each of the five $\mathcal N=4$ vector multiplets decomposes into an $\mathcal N=2$ vector multiplet with fields $(F_{\mu\nu}^{(+)\,a},\chi^{(+)\,a},P_\mu^{(+-)\,a9})$ and an $\mathcal N=2$ hypermultiplet with fields $(P_\mu^{(+-)\,a\hat b},\chi^{(-)\,a})$.   Here the $(\pm)$ superscripts on the fermions indicate their $t^9$ eigenvalues, and $\tilde\psi_\mu^{(+)}=\psi_\mu^{(+)}-\fft{i}3\gamma_\mu\lambda^{(+)}$ is the shifted gravitino.  By singling out the $y^9$ direction, we have decomposed the vector of $Sp(4)$ as $5\to4+1$, so the hatted indices $\hat a$ and $\hat b$ run from $1,\cdots,4$.  This decomposition can be seen more explicitly in the leading order supersymmetry variations obtained from (\ref{eq:redsusy})
\begin{align}
    \mbox{supergravity}:&&\delta_\epsilon\psi_\mu^{(+)}&=\left[\nabla_\mu-\fft18h_{\mu\nu\rho}\gamma^{\nu\rho}-\fft14Q_\mu^{(--)\,\hat a\hat b}t^{\hat a\hat b}+\fft{i}4F_{\mu\nu}^{(-)\,9}\gamma^\nu\right]\epsilon,\nn\\
    \mbox{gravitino}:&&\delta_\epsilon\psi^{(-)}_\mu&=\left[-\fft12Q_\mu^{(--)\,\hat a5}t^{\hat a}+\fft{i}4F_{\mu\nu}^{(-)\,\hat a}\gamma^\nu t^{\hat a}\right]\epsilon,\nn\\
    &&\delta_\epsilon\lambda^{(-)}&=\left[\fft18F_{\mu\nu}^{(-)\,\hat a}\gamma^{\mu\nu}t^{\hat a}\right]\epsilon,\nn\\
    \mbox{vector}:&&\delta_\epsilon\lambda^{(+)}&=\left[-i\gamma^\mu\partial_\mu\varphi+\fft{i}{12}h_{\mu\nu\rho}\gamma^{\mu\nu\rho}+\fft18F_{\mu\nu}^{(-)\,9}\gamma^{\mu\nu}\right]\epsilon,\nn\\
    \mbox{vector}:&&\delta_\epsilon\chi^{(+)\,a}&=\left[-\fft{i}2\gamma^\mu P_\mu^{(+-)\,a5}-\fft18F_{\mu\nu}^{(+)\,a}\gamma^{\mu\nu}\right]\epsilon,\nn\\
    \mbox{hyper}:&&\delta_\epsilon\chi^{(-)\,a}&=\left[-\fft{i}2\gamma^\mu P_\mu^{(+-)\,a\hat b}t^{\hat b}\right]\epsilon.
\end{align}
Note that these transformations are given in the string frame, while the standard supergravity variations are usually presented in the Einstein frame and with the three-form $h$ dualized to a two-form field strength.  Nevertheless, the truncation to an $\mathcal N=2$ subsector is clear even in the string frame.

We can truncate the $\mathcal N=4$ theory to $\mathcal N=2$ supergravity coupled to six vector multiplets by removing the gravitino multiplet and hypermultiplets.  This can be accomplished by setting
\begin{equation}
    e_i{}^a=\begin{pmatrix}\mathbf1&\mathbf0\\\phi_a&e^\sigma\end{pmatrix},\qquad
    b_{ij}=\begin{pmatrix}\mathbf0&-\phi_i\\\phi_j&0\end{pmatrix},\qquad
    A_\mu^i=B_{\mu i},
\end{equation}
so that
\begin{align}
    &Q_\mu^{(--)\,ab}=0,\qquad P_\mu^{(+-)\,\hat a5}=e^{-\sigma}\partial_\mu\phi_{\hat a},\qquad P_\mu^{(+-)\,55}=\partial_\mu\sigma,\nn\\
    &F_{\mu\nu}^{(+)\,\hat a}=2(F_{\mu\nu}^i+\phi_iF_{\mu\nu}^5),\qquad F_{\mu\nu}^{(-)\,\hat a}=0,\nn\\
    &F_{\mu\nu}^{(\pm)\,5}=e^\sigma F^5\pm e^{-\sigma}(G_{\mu\nu\,5}-2\phi_iF_{\mu\nu}^i-(\phi_i)^2F_{\mu\nu}^5).
\end{align}
The resulting $\mathcal N=2$ supersymmetry variations then take the form
\begin{align}
    \delta_\epsilon\psi_\mu^{(+)}&=\Bigl[\nabla_\mu(\tilde\omega_-)+\fft{i}4F_{\mu\nu}^{(-)\,5}(1-\alpha'T)\gamma^\nu\nn\\
    &\quad+\fft{i\alpha'}8\Bigl(\fft12R_{\mu\nu\alpha\beta}(\tilde\omega_+)F_{\alpha\beta}^{(-)\,5}+2e^{-\sigma}\partial_\alpha\phi_{\hat a}\nabla_{[\mu}^{\prime(+)}F_{\nu]\alpha}^{(+)\,\hat a}+2\partial_\alpha\sigma\nabla_{[\mu}^{\prime(+)}F_{\nu]\alpha}^{(+)\,5}\nn\\
    &\kern4em-\fft14F_{\alpha\beta}^{(-)\,5}F_{\beta[\mu}^{(+)\,\hat a}F_{\nu]\alpha}^{(+)\,\hat a}-\fft14F_{\alpha\beta}^{(-)\,5}F_{\beta[\mu}^{(+)\,5}F_{\nu]\alpha}^{(+)\,5}\Bigr)\gamma^\nu\Bigr]\epsilon,\nn\\
    \delta_\epsilon\lambda^{(+)}&=\Bigl[-i\gamma^\mu\partial_\mu\varphi+\fft{i}{12}\tilde h_{\mu\nu\rho}\gamma^{\mu\nu\rho}+\fft18F_{\mu\nu}^{(-)\,5}(1-\alpha'T)\gamma^{\mu\nu}\nn\\
    &\quad+\fft{\alpha'}{16}\Bigl(\fft12R_{\mu\nu\alpha\beta}(\tilde\omega_+)F_{\alpha\beta}^{(-)\,5}+2e^{-\sigma}\partial_\alpha\phi_{\hat a}\nabla_{[\mu}^{\prime(+)}F_{\nu]\alpha}^{(+)\,\hat a}+2\partial_\alpha\sigma\nabla_{[\mu}^{\prime(+)}F_{\nu]\alpha}^{(+)\,5}\nn\\
    &\kern4em-\fft14F_{\alpha\beta}^{(-)\,5}F_{\beta[\mu}^{(+)\,\hat a}F_{\nu]\alpha}^{(+)\,\hat a}-\fft14F_{\alpha\beta}^{(-)\,5}F_{\beta[\mu}^{(+)\,5}F_{\nu]\alpha}^{(+)\,5}\Bigr)\gamma^{\mu\nu}\Bigr]\epsilon,\nn\\
    \delta_\epsilon\chi^{(+)\,\hat a}&=\left[-\fft{i}2e^{-\sigma}\gamma^\mu \partial_\mu\phi_{\hat a}(1-\alpha'T)-\fft18F_{\mu\nu}^{(+)\,\hat a}\gamma^{\mu\nu}\right]\epsilon,\nn\\
    \delta_\epsilon\chi^{(+)\,5}&=\left[-\fft{i}2\gamma^\mu\partial_\mu\sigma(1-\alpha'T)-\fft18F_{\mu\nu}^{(+)\,5}\gamma^{\mu\nu}\right]\epsilon.
\end{align}
The structure of these variations suggests that much of the complication arising from the $\mathcal O(\alpha')$ corrections is related to the $F_{\mu\nu}^{(-)\,5}$ component of the graviphoton.

A common further truncation is to remove the four vector multiplets $(F_{\mu\nu}^{(+)\,\hat a},\chi^{(+)\,\hat a},\partial_\mu\phi_{\hat a})$ related to the $T^4$ that we are not concerned with.  The final field content is that of the five-dimensional STU model (presented here in the string frame and with the three-form $h$), with bosonic Lagrangian given in (\ref{eq:4derlag}) and with corresponding fermionic supersymmetry variations
\begin{align}
    \delta_\epsilon\psi_\mu^{(+)}&=\Bigl[\nabla_\mu(\tilde\omega_-)+\fft{i}4F_{\mu\nu}^{(-)}(1-\alpha'T)\gamma^\nu\nn\\
    &\quad+\fft{i\alpha'}8\Bigl(\fft12R_{\mu\nu\alpha\beta}(\tilde\omega_+)F_{\alpha\beta}^{(-)}+2\partial_\alpha\sigma\nabla_{[\mu}^{\prime(+)}F_{\nu]\alpha}^{(+)}-\fft14F_{\alpha\beta}^{(-)}F_{\beta[\mu}^{(+)}F_{\nu]\alpha}^{(+)}\Bigr)\gamma^\nu\Bigr]\epsilon,\nn\\
    \delta_\epsilon\lambda^{(+)}&=\Bigl[-i\gamma^\mu\partial_\mu\varphi+\fft{i}{12}\tilde h_{\mu\nu\rho}\gamma^{\mu\nu\rho}+\fft18F_{\mu\nu}^{(-)}(1-\alpha'T)\gamma^{\mu\nu}\nn\\
    &\quad+\fft{\alpha'}{16}\Bigl(\fft12R_{\mu\nu\alpha\beta}(\tilde\omega_+)F_{\alpha\beta}^{(-)}+2\partial_\alpha\sigma\nabla_{[\mu}^{\prime(+)}F_{\nu]\alpha}^{(+)}-\fft14F_{\alpha\beta}^{(-)}F_{\beta[\mu}^{(+)}F_{\nu]\alpha}^{(+)}\Bigr)\gamma^{\mu\nu}\Bigr]\epsilon,\nn\\
    \delta_\epsilon\chi^{(+)}&=\left[-\fft{i}2\gamma^\mu\partial_\mu\sigma(1-\alpha'T)-\fft18F_{\mu\nu}^{(+)}\gamma^{\mu\nu}\right]\epsilon.
\end{align}
Here we have dropped the index $5$ as there is no ambiguity in the compactification on a single circle.

We now return to the supersymmetry variations of the three-charge BPS black hole solution, given above in (\ref{eq:3Qsusy}) in ten-dimensional Dirac notation.  When reduced to five-dimensional $\mathcal N=2$ notation, the resulting supersymmetry variations take the form
\begin{align}
    \delta_\epsilon\psi^{(+)}_t&=\gamma_t{}^i\partial_ifP_+\epsilon,\nn\\
    \delta_\epsilon\psi^{(+)}_i&=e^{f/2}\qty[\partial_i+(\partial_if+\gamma_i{}^j\partial_jg)P_+]e^{-f/2}\epsilon,\nn\\
    \delta_\epsilon\lambda^{(+)}&=-2i\gamma^i\partial_i\varphi P_+\epsilon,\nn\\
    \delta_\epsilon\chi^{(+)}&=-i\gamma^i\partial_i\sigma(1-\alpha'T)P_+\epsilon,
\end{align}
where the 1/2-BPS projection is given by
\begin{equation}
    P_\pm=\fft12(1\pm i\gamma^0).
\end{equation}
The $\mathcal N=2$ killing spinor then takes the form $\epsilon=e^{f/2}P_-\epsilon_0$, where $\epsilon_0$ is a constant spinor.

\subsection{Field-redefined forms of the solution}

Although the dimensionally reduced heterotic action is naturally written in terms of the three-form field strength $h$, we may dualize $h$ into a two-form field strength in order to facilitate comparison with canonical five-dimensional $\mathcal N=2$ supergravity coupled to two vector multiplets.  Before doing so,   however, we present the field-redefined three-charge black hole solution corresponding to the Lagrangians (\ref{eq:4derlag2}) and (\ref{eq:4derlag3}).

The original form of the reduced heterotic Lagrangian, (\ref{eq:4derlag}), involves derivatives of field strengths such as $(\nabla F)^2$.  Such terms can be removed by field redefinition, yielding the Lagrangian, (\ref{eq:4derlag2}).  These field redefinitions transform the original three-charge black hole solution, (\ref{eq:3chargesol}), into the form
\begin{align}
    \dd s^2&=-\fft1{H_1H_2}\qty(1-\alpha'\fft{(\partial_i\log(H_1/H_2))^2}{4H_3})\dd t^2\nn\\
    &\qquad+H_3\qty(1+\alpha'\fft{(\partial_i\log H_3)^2}{4H_3})\qty(\dd x^i\dd x^i-\fft{\alpha'}{8H_3}\left(\partial_i\log(H_1H_2)\dd x^i\right)^2),\nn\\
    \varphi&=\fft14\log(\fft{(H_3)^2}{H_1H_2})-\alpha'\fft{(\partial_i\log(H_1H_2))^2+2(\partial_i\log(H_1/H_2))^2-4(\partial_i\log H_3)^2}{32H_3},\nn\\
    \sigma&=\fft12\log(\fft{H_1}{H_2}),\nn\\
    A_t&=\fft1{H_1}\qty(1+\alpha'\fft{(\partial_i\log(H_1H_2))^2}{8H_3}),\nn\\
    B_t&=-\fft1{H_2}\qty(1+\alpha'\fft{(\partial_i\log(H_1H_2))^2}{8H_3}),\nn\\
    h_{ijk}&=\epsilon_{ijkl}\partial_lH_3.
\end{align}
Note that this reduces to the two-charge solution, (\ref{eq:fr2cbh}), in the limit when the three-form charge is turned off (\textit{i.e.}~when $H_3=1$).

We have also considered the field-redefinition framework involving a torsion-free connection.  In this case, the Lagrangian is given by (\ref{eq:4derlag3}), and the three-charge black hole takes the form:

\begin{align}
    \dd s^2&=-\fft1{H_1H_2}\qty(1-\alpha'\fft{(\partial_i\log(H_1/H_2))^2}{4H_3})\dd t^2\nn\\
    &\qquad+H_3\qty(1+\alpha'\fft{(\partial_i\log H_3)^2}{2H_3})\qty(\dd x^i\dd x^i-\alpha'\fft{\left(\partial_i\log(H_1H_2)\dd x^i\right)^2 + 2\left(\partial_i\log(H_3)\dd x^i\right)^2}{8H_3}),\nn\\
    \varphi&=\fft14\log(\fft{(H_3)^2}{H_1H_2})-\alpha'\fft{(\partial_i\log(H_1H_2))^2+2(\partial_i\log(H_1/H_2))^2-10(\partial_i\log H_3)^2}{32H_3},\nn\\
    \sigma&=\fft12\log(\fft{H_1}{H_2}),\nn\\
    A_t&=\fft1{H_1}\qty(1+\alpha'\fft{(\partial_i\log(H_1H_2))^2}{8H_3}),\nn\\
    B_t&=-\fft1{H_2}\qty(1+\alpha'\fft{(\partial_i\log(H_1H_2))^2}{8H_3}),\nn\\
    h_{ijk}&=\epsilon_{ijkl}\partial_lH_3.
\end{align}

\subsection{Dualization of the heterotic action}

To facilitate comparison with the standard formulation of five-dimensional supergravity coupled to vector multiplets, we may dualize the heterotic three-form $h$.  In particular, we dualize the final field-redefined Lagrangian, (\ref{eq:4derlag3}), which is written using a torsion-free connection.  In this case, the three-form $h$ only appears explicitly in the Lagrangian, so dualization is straightforward.

Dualizing $h$ exchanges its Bianchi identity with its equation of motion, where
\begin{align}
    \dd\tilde h &= \dfrac{1}{4} F^{(-)} \wedge F^{(-)} - \dfrac{1}{4} F^{(+)} \wedge F^{(+)} - \dfrac{\alpha'}{4} \Tr[R \wedge R],\nn\\
    \dd(e^{-2\varphi}*\tilde h)&=0.
\end{align}
This suggests that we introduce a two-form field strength $H=\dd C$ such that
\begin{equation}
    \tilde h=e^{2\varphi}*H.
\end{equation}
We proceed to dualize the Lagrangian, (\ref{eq:4derlag3}), by introducing a one-form Lagrange multiplier $C$ and adding a term
\begin{equation}
    \Delta \mathcal{L} = C \wedge \qty(\dd \tilde{h} -\dfrac{1}{4} F^{(-)} \wedge F^{(-)} + \dfrac{1}{4} F^{(+)} \wedge F^{(+)} + \dfrac{\alpha'}{4} \Tr\qty[R \wedge R]).
\end{equation}
Varying with respect to $C$ imposes the Bianchi identity on $\tilde h$. However, varying with respect to $\tilde{h}$ gives
\begin{equation}
    \tilde{h}_{\alpha\beta\gamma} = \dfrac{1}{2} \,  e^{2\varphi}\,\epsilon_{\mu\nu\alpha\beta\gamma}\, H^{\mu\nu} + \fft{\alpha'}8 \dfrac{\delta \mathcal{L}_{\partial^4}}{\delta h^{\alpha\beta\gamma}},
\label{eq:tildehdual}
\end{equation}
where $\mathcal{L}_{\partial^4}$ is the four-derivative part of the Lagrangian.  Substituting $\tilde h$ back into the Lagrangian then gives
\begin{align}
    \mathcal{L}\,\dd[5]x &= \sqrt{-g}e^{-2\varphi} \qty(R + 4(\partial\varphi)^2 - (\partial \sigma)^2 - \dfrac{1}{8} \qty(F^{(-)\,2} + F^{(+)\,2}) - \dfrac{1}{4} e^{4\varphi} H^2)\,\dd[5]x \nn\\
    &\quad+ \dfrac{1}{4} C \wedge \left( F^{(+)} \wedge F^{(+)} - F^{(-)} \wedge F^{(-)}+\alpha' \Tr[R \wedge R]\right)\nn\\
    &\quad+\fft{\alpha'}8 \mathcal{L}_{\partial^4}\,\dd[5]x,
\label{eq:finalLag}
\end{align}
where $\mathcal L_{\partial^4}$ is simply the four-derivative part of (\ref{eq:4derlag3}) with $\tilde h$ replaced by $H$.  In particular, the contribution of the $\mathcal O(\alpha')$ variation in (\ref{eq:tildehdual}) cancels when substituting into the two-derivative Lagrangian, and hence its explicit form is unimportant.

Before writing out $\mathcal L_{\partial^4}$, we note that, after dualization of $h$, the two derivative equations of motion take the form
\begin{align}
    \mathcal{E}_{g,\alpha\beta}\kern.7em& = R_{\alpha\beta} - \dfrac{1}{4}\left(F^{(+)\,2}_{\alpha\beta} + F^{(-)\,2}_{\alpha\beta}\right) - \dfrac{1}{2} e^{4\varphi} \Bigl(H^2_{\alpha\beta}-\fft12 H^2 g_{\alpha\beta}\Bigr)+ 2\nabla_\alpha \nabla_\beta \varphi  - \partial_\alpha\sigma \partial_\beta\sigma,\nn\\
    \mathcal{E}_{F^{(-)},\alpha} &= e^{2\varphi}\nabla^\beta\qty(e^{-2\varphi}F^{(-)}_{\alpha\beta}) + F^{(+)}_{\alpha\beta} \partial^\beta\sigma - \dfrac{1}{4} e^{2\varphi} \epsilon_\alpha{}^{\beta\gamma\delta\epsilon} H_{\beta\gamma} F^{(-)}_{\delta\epsilon},\nn \\
    \mathcal{E}_{F^{(+)},\alpha} &=e^{2\varphi}\nabla^\beta\qty(e^{-2\varphi} F^{(+)}_{\alpha\beta})  + F^{(-)}_{\alpha\beta} \partial^\beta\sigma + \dfrac{1}{4} e^{2\varphi} \epsilon_\alpha{}^{\beta\gamma\delta\epsilon} H_{\beta\gamma} F^{(+)}_{\delta\epsilon},\nn \\
    \mathcal{E}_{H,\alpha}\kern.9em&=e^{-2\varphi}\nabla^{\beta}\qty(e^{2\varphi}H_{\alpha\beta})  + \dfrac{1}{16}e^{-2\varphi} \epsilon_\alpha{}^{\beta\gamma\delta\epsilon} \left(F^{(-)}_{\beta\gamma} F^{(-)}_{\delta\epsilon} - F^{(+)}_{\alpha\beta} F^{(+)}_{\delta\epsilon}\right),\nn\\
    \mathcal{E}_\sigma\kern1.8em&=\Box \sigma - 2\partial_\alpha\sigma\partial^\alpha\varphi - \dfrac{1}{4} F^{(-)}_{\alpha\beta} F^{(+)\,\alpha\beta},\nn\\
    \mathcal{E}_\varphi\kern1.8em & =\Box\varphi + \dfrac{1}{4} R - \dfrac{1}{32} \left( F^{(-)\,2} +  F^{(+)\,2}\right) + \dfrac{1}{16} e^{4\varphi} H^2 - (\partial\varphi)^2 - \dfrac{1}{4} (\partial\sigma)^2,\label{eq: EOM dualised}
\end{align}
where we have defined $H^2_{\alpha\beta}=H_{\alpha\gamma}H_\beta{}^\gamma$, and likewise for $F^{(-)\,2}_{\alpha\beta}$ and $F^{(+)\,2}_{\alpha\beta}$. Dualization of $h$ in $\mathcal  L_{\partial^4}$ is straightforward.   However, note that dualization of $Rh^2$ gives rise to
\begin{equation}
    R_{\alpha\beta\gamma\delta} h^{\alpha\beta\epsilon} h^{\gamma\delta}{}_{\epsilon} = -e^{4\varphi} \qty(R_{\alpha\beta\gamma\delta} H^{\alpha\beta} H^{\gamma\delta} - 4 R^{\alpha\beta} H^2_{\alpha\beta} +RH^2 ).
\end{equation}
Since we are only working up to $\mathcal O(\alpha')$ throughout, we may remove the Ricci terms by field redefinitions.  Using the equations of motion, \eqref{eq: EOM dualised}, we find
\begin{align}
     - 4R^{\alpha\beta}H^2_{\alpha\beta} +RH^2= &\;\; \dfrac{1}{4} \qty(F^{(-)\,2} + F^{(+)\,2})H^2 - \qty(F^{(-)\,2}_{\alpha\beta} + F^{(+)\,2}_{\alpha\beta})H^2_{\alpha\beta} \nn \\
    &+ \dfrac{1}{4}e^{4\varphi} (H^2)^2 - 2e^{4\varphi}H^2_{\alpha\beta} H^2_{\alpha\beta} + \qty((\partial\sigma)^2 + 4(\partial\varphi)^2)H^2 \nn \\
    &-4 H^2_{\alpha\beta}\partial^\alpha\sigma\partial^\beta\sigma - \dfrac{1}{2}e^{-2\varphi}\epsilon^{\alpha\beta\gamma\delta\epsilon} H_{\eta\epsilon}\partial^\eta\varphi \qty(F^{(-)}_{\alpha\beta} F^{(-)}_{\gamma\delta} - F^{(+)}_{\alpha\beta} F^{(+)}_{\gamma\delta})\nn \\
    &+ \qty(H^2 g_{\alpha\beta} - 4H^2_{\alpha\beta})\mathcal{E}_{g,\alpha\beta} + 8 H^{\beta\alpha}\partial_\beta\varphi \mathcal{E}_{H,\alpha}\nn\\
    &+e^{-2\varphi}\nabla^\mu\qty(e^{2\varphi}\qty(4H^2_{\mu\nu}\partial^\nu\varphi-H^2\partial_\mu\varphi)).
\end{align}
The equation of motion terms here can be removed by the field redefinitions
\begin{align}
    \delta g_{\mu\nu} &= \dfrac{1}{2}e^{4\varphi} \qty( 4 H^2_{\mu\nu}-H^2 g_{\mu\nu} ),\nn\\
    \delta\varphi &= -\dfrac{1}{8} e^{4\varphi} H^2,\nn\\
    \delta C_\mu &= -4 H_{\mu\alpha}\partial^\alpha\varphi.
\end{align}
Then, the dualized Lagrangian takes the form of (\ref{eq:finalLag}) with
\begin{align}
    e^{-1}\mathcal{L}_{\partial^4} =&\; e^{-2\varphi} \Biggl[\qty(R_{\alpha\beta\gamma\delta})^2 - \dfrac{1}{2} R^{\alpha\beta\gamma\delta} F^{(-)}_{\alpha\beta} F^{(-)}_{\gamma\delta} + \dfrac{1}{32} \Big(F^{(-)\,2} F^{(-)\,2} - F^{(+)\,2} F^{(+)\,2}\Big)\nn \\
    &\qquad - \dfrac{1}{16} \Big(F^{(-)\,2}_{\alpha\beta} F^{(-)\,2}_{\alpha\beta} + F^{(+)\,2}_{\alpha\beta} F^{(+)\,2}_{\alpha\beta} \Big) - \dfrac{1}{4} F^{(-)\,2}_{\alpha\beta} F^{(+)\,2}_{\alpha\beta}\nn \\
    &\qquad + \dfrac{1}{2}  \Big(F^{(-)\,2} - F^{(+)\,2}\Big)(\partial\sigma)^2 + \qty(F^{(+)\,2}_{\alpha\beta} - F^{(-)\,2}_{\alpha\beta})\partial^\alpha\sigma\partial^\beta\sigma \Biggr]\nn \\
    & + e^{2\varphi} \Biggl[ \dfrac{1}{2}  R^{\alpha\beta\gamma\delta} H_{\alpha\beta} H_{\gamma\delta} + \dfrac{1}{8} H^2 \Big(2 F^{(-)\,2} + 3 F^{(+)\,2} \Big)\nn \\
    &\qquad - \dfrac{1}{2} H^2_{\alpha\beta} \Big(F^{(-)\,2}_{\alpha\beta} + 2F^{(+)\,2}_{\alpha\beta} \Big) + \dfrac{1}{16} H^{\alpha\beta} H^{\gamma\delta} \Big( F^{(+)}_{\alpha\beta} F^{(+)}_{\gamma\delta} - F^{(-)}_{\alpha\beta} F^{(-)}_{\gamma\delta} \Big) \nn \\
    &\qquad -\dfrac{1}{8} H^{\alpha\beta}H^{\gamma\delta} \qty(F^{(-)}_{\beta\gamma}  F^{(-)}_{\delta\alpha}  +F^{(+)}_{\beta\gamma} F^{(+)}_{\delta\alpha}) -4 H^2_{\alpha\beta} \partial^\alpha\sigma\partial^\beta\sigma \nn \\
    &\qquad + \dfrac{1}{2} H^2 \Big(3 (\partial\sigma)^2 + 4(\partial\varphi)^2\Big) \Biggr]-\fft{11}8e^{6\varphi}H^2_{\alpha\beta}H^2_{\alpha\beta}\nn\\
    &+ \dfrac{1}{4} \epsilon^{\mu\nu\alpha\beta\gamma} H_{\mu\nu} \Big( F^{(+)}_{\alpha\delta} F^{(-)}_{\beta\gamma} \partial^\delta\sigma + F^{(-)}_{\alpha\delta} F^{(+)}_{\beta\gamma}\partial^\delta\sigma -2 F^{(-)}_{\alpha\delta} F^{(+)}_{\gamma\delta} \partial_\beta\sigma\Big)\nn \\
    &- \dfrac{1}{4} \epsilon^{\mu\nu\rho\sigma\alpha} H_{\beta\alpha}  \qty( F^{(-)}_{\mu\nu} F^{(-)}_{\rho\sigma} - F^{(+)}_{\mu\nu} F^{(+)}_{\rho\sigma}) \partial^\beta \varphi.
\label{eq:final4d}
\end{align}

The dualized Lagrangian, (\ref{eq:finalLag}) with (\ref{eq:final4d}), is still written in the string frame.  To transform to the Einstein frame, we perform the following Weyl scaling of the metric
\begin{equation}
    g_{\mu\nu} \to e^{4\varphi/3}g_{\mu\nu}.
\end{equation}
This gives
\begin{align}
        \mathcal{L}\,\dd[5]x =&\;\;\sqrt{-g}\left( R - \dfrac{4}{3} (\partial\varphi)^2 - (\partial\sigma)^2 - \dfrac{1}{8}e^{-4\varphi/3} (F^{(-)\,2} + F^{(+)\,2}) - \dfrac{1}{4} e^{8\varphi/3} H^2\right)\dd[5]x \nonumber\\
        &+ \dfrac{1}{4} C \wedge \qty(F^{(+)}\wedge F^{(+)} - F^{(-)} \wedge F^{(-)}+\alpha' \Tr[R \wedge R]) + \dfrac{\alpha'}{8} \, \mathcal{L}_{\partial^4}\,\dd[5]x,
\end{align}
where the four-derivative part takes the form,
\begin{align}
    e^{-1}\mathcal{L}_{\partial^4} =&\ e^{-4\varphi/3}\tilde R_{\alpha\beta\gamma\delta}^2 - e^{-8\varphi/3}\Bigl[ \dfrac{1}{2}\tilde R^{\alpha\beta\gamma\delta}F^{(-)}_{\alpha\beta} F^{(-)}_{\gamma\delta} - \dfrac{1}{2}  \Big(F^{(-)\,2} - F^{(+)\,2}\Big)(\partial\sigma)^2 \nn\\
    &+ \qty(F^{(-)\,2}_{\alpha\beta} - F^{(+)\,2}_{\alpha\beta})\partial^\alpha\sigma\partial^\beta\sigma\Bigr] \nn\\
    & + e^{-4\varphi} \Bigl[\dfrac{1}{32} \Big(F^{(-)\,2}F^{(-)\,2} - F^{(+)\,2}F^{(+)\,2}\Big) - \dfrac{1}{4} F^{(-)\,2}_{\alpha\beta} F^{(+)\,2}_{\alpha\beta} \nn\\
    &- \dfrac{1}{16} \Big(F^{(-)\,2}_{\alpha\beta}F^{(-)\,2}_{\alpha\beta} + F^{(+)\,2}_{\alpha\beta}F^{(+)\,2}_{\alpha\beta}\Big)\Bigr] - \dfrac{11}{8} e^{4\varphi} H^2_{\alpha\beta} H^2_{\alpha\beta}\nn\\
    &+ e^{4\varphi/3} \Bigl[ \dfrac{1}{2} \tilde R^{\alpha\beta\gamma\delta}H_{\alpha\beta} H_{\gamma\delta} -4 H^2_{\alpha\beta} \partial^\alpha\sigma\partial^\beta\sigma  + \dfrac{1}{2}H^2 \Big(3(\partial\sigma)^2 + 4\partial(\varphi)^2\Big) \Bigr]\nn\\
    & - \dfrac{1}{2} H^2_{\alpha\beta} \Big(F^{(-)\,2}_{\alpha\beta} + 2F^{(+)\,2}_{\alpha\beta} \Big) + \dfrac{1}{16} H^{\alpha\beta} H^{\gamma\delta} \Big( F^{(+)}_{\alpha\beta} F^{(+)}_{\gamma\delta} - F^{(-)}_{\alpha\beta} F^{(-)}_{\gamma\delta} \Big) \nn \\
    & -\dfrac{1}{8} H^{\alpha\beta}H^{\gamma\delta} \qty(F^{(-)}_{\beta\gamma} F^{(-)}_{\delta\alpha}+F^{(+)}_{\beta\gamma}  F^{(+)}_{\delta\alpha}) + \dfrac{1}{8}\, H^2 \Big(2F^{(-)\, 2} + 3F^{(+)\,2}\Big)\nn\\
    &+e^{-4\varphi/3} \Bigl[ \dfrac{1}{4} \epsilon^{\mu\nu\alpha\beta\gamma} H_{\mu\nu} \Big( F^{(+)}_{\alpha\delta} F^{(-)}_{\beta\gamma} \partial^\delta\sigma + F^{(-)}_{\alpha\delta} F^{(+)}_{\beta\gamma}\partial^\delta\sigma -2 F^{(-)}_{\alpha\delta} F^{(+)}_{\gamma\delta} \partial_\beta\sigma\Big)\nn \\
    &- \dfrac{1}{4} \epsilon^{\mu\nu\rho\sigma\alpha} H_{\beta\alpha}  \Big( F^{(-)}_{\mu\nu} F^{(-)}_{\rho\sigma} - F^{(+)}_{\mu\nu} F^{(+)}_{\rho\sigma} \Big) \partial^\beta \varphi\Bigr].
\end{align}
Here, $\tilde R_{\alpha\beta\gamma\delta}$ is the Weyl-scaled Riemann tensor, and is given explicitly by
\begin{align}
    \tilde R^\alpha{}_{\beta\gamma\delta} &= R^\alpha{}_{\beta\gamma\delta} - \dfrac{4}{3} \Big(\delta^\alpha_{[\gamma} \delta^\mu_{\delta]}\delta^\nu_\beta - g\vphantom{\delta}_{\beta[\gamma}^{\vphantom{\alpha}}\delta^\mu_{\delta]}g^{\alpha\nu}\Big) \nabla_\mu\nabla_\nu\varphi\nn\\
    &\quad+ \dfrac{8}{9}\Big(\delta^\alpha_{[\gamma} \delta^\mu_{\delta]}\delta^\nu_\beta - g\vphantom{\delta}^{\vphantom{\alpha}}_{\beta[\gamma}\delta^\mu_{\delta]}g^{\alpha\nu} + g\vphantom{\delta}^{\vphantom{\alpha}}_{\beta[\gamma}\delta^\alpha_{\delta]} g^{\mu\nu} \Big)\partial_\mu\varphi\,\partial_\nu\varphi.
\end{align}
Note that the Weyl scaling has introduced second derivatives of the dilaton, namely terms containing $\nabla_\mu\nabla_\nu\varphi$.  These terms can be removed by integration by parts along with another set of field redefinitions
\begin{align}
    \delta g_{\mu\nu} &= \frac{80}{9}e^{-4\varphi/3} \left(\partial_\mu \varphi\, \partial_\nu \varphi-\frac{1}{5}(\partial\varphi)^2g_{\mu\nu} \right) + \dfrac{1}{9} e^{-8\varphi/3} \qty(F^{(+)\, 2}+F^{(-)\, 2}) g_{\mu\nu} - \dfrac{4}{9} e^{4\varphi/3} H^2 g_{\mu\nu},\nn\\
    \delta A_\mu &= \dfrac{4}{3} e^{-4\varphi/3-\sigma} F_{\mu\nu}^{(-)} \partial^\nu\varphi,\nn\\
    \delta B_\mu &= -\dfrac{4}{3} e^{-4\varphi/3+\sigma} F_{\mu\nu}^{(-)} \partial^\nu\varphi,\nn\\
    \delta C_\mu &= -\dfrac{4}{3} e^{-4\varphi/3} H_{\mu\nu} \partial^\nu \varphi,\nn\\
    \delta\varphi&= - \dfrac{3}{8} e^{-8\varphi/3} \qty(F^{(+)\, 2}+F^{(-)\, 2}) + \dfrac{9}{8} e^{4\varphi/3}H^2 - e^{-4\varphi/3}\left(\dfrac{12}{9} (\partial\varphi)^2 + (\partial\sigma)^2\right).
\end{align}
After doing this, we end up with the final form of the $\alpha'$-corrected Lagrangian
\begin{align}
    \mathcal{L}\,\dd[5]x =&\;\;\sqrt{-g}\left( R - \dfrac{4}{3} (\partial\varphi)^2 - (\partial\sigma)^2 - \dfrac{1}{8}(\tilde F^{(-)\,2} + \tilde F^{(+)\,2}) - \dfrac{1}{4}\tilde H^2 \right)\dd[5]x\nonumber\\
    &+ \dfrac{1}{4} C \wedge \qty(F^{(+)}\wedge F^{(+)} - F^{(-)} \wedge F^{(-)}+\alpha' \Tr[R \wedge R]) + \dfrac{\alpha'}{8} \, \mathcal{L}_{\partial^4}\,\dd[5]x,
\label{eq:finalL2}
\end{align}
where
\begin{align}
    e^{-1}\mathcal L_{\partial^4}&=\,e^{-4\varphi/3}\biggl[R_{\alpha\beta\gamma\delta}^2+\fft12R^{\alpha\beta\gamma\delta}\qty(\tilde H_{\alpha\beta}\tilde H_{\gamma\delta}-\tilde F^{(-)}_{\alpha\beta}\tilde F^{(-)}_{\gamma\delta})-\fft{128}{27}\qty((\partial\varphi)^2)^2-\fft{80}9(\partial\varphi\cdot\partial\sigma)^2\nn\\
    &\kern5em+\fft{16}9(\partial\varphi)^2(\partial\sigma)^2-\fft49\left(5\tilde F^{(+)\,2}_{\mu\nu}+2\tilde F^{(-)\,2}_{\mu\nu}+12\tilde H^2_{\mu\nu}\right)\partial^\mu\varphi\partial^\nu\varphi\nn\\
    &\kern5em
    +\left(\tilde F^{(+)\,2}_{\mu\nu}-\tilde F^{(-)\,2}_{\mu\nu}-4\tilde H^2_{\mu\nu}\right)\partial^\mu\sigma\partial^\nu\sigma+\fft43\tilde F^{(+)}_{\mu\lambda}\tilde F^{(-)}_{\nu\lambda}\qty(\partial^\mu\sigma\partial^\nu\varphi-\partial^\mu\varphi\partial^\nu\sigma)\nn\\
    &\kern5em+\fft29\qty(3\tilde F^{(+)\,2}+\tilde F^{(-)\,^2}+5\tilde H^2)(\partial\varphi)^2+\fft16\qty(-2\tilde F^{(+)\,2}+4\tilde F^{(-)\,^2}+5\tilde H^2)(\partial\sigma)^2\nn\\
    &\kern5em+\fft23\tilde F^{(+)}_{\mu\nu}\tilde F^{(-)}_{\mu\nu}(\partial\varphi\cdot\partial\sigma)+\fft1{288}\qty(\tilde F^{(+)\,2}+\tilde F^{(-)\,2})\qty(\tilde F^{(+)\,2}+13\tilde F^{(-)\,2})\nn\\
    &\kern5em+\fft1{144}\tilde H^2\qty(23\tilde F^{(+)\,2}+17\tilde F^{(-)\,^2}+44\tilde H^2)+\fft1{16}\left((\tilde H_{\mu\nu}\tilde F^{(+)}_{\mu\nu})^2-(\tilde H_{\mu\nu}\tilde F^{(-)}_{\mu\nu})^2\right)\nn\\
    &\kern5em-\fft1{16}\qty(\tilde F^{(+)\,4}+\tilde F^{(-)\,4})-\fft14\tilde F^{(+)}\tilde F^{(+)}\tilde F^{(-)}\tilde F^{(-)}-\fft18\tilde H\tilde F^{(+)}\tilde H\tilde F^{(+)}\nn\\
    &\kern5em-\fft18\tilde H\tilde F^{(-)}\tilde H\tilde F^{(-)}-\tilde H\tilde H\tilde F^{(+)}\tilde F^{(+)}-\fft12\tilde H\tilde H\tilde F^{(-)}\tilde F^{(-)}-\fft{11}8\tilde H^4\nn\\
    &\kern5em-\fft16\epsilon^{\mu\nu\rho\sigma\alpha}\tilde F^{(+)}_{\mu\nu}\tilde F^{(+)}_{\rho\sigma}\tilde H_{\alpha\beta}\partial^\beta\varphi-\fft14\epsilon^{\mu\nu\rho\sigma\alpha}\tilde F^{(+)}_{\mu\nu}\tilde F^{(-)}_{\rho\sigma}\tilde H_{\alpha\beta}\partial^\beta\sigma\nn\\
    &\kern5em+\fft12\epsilon^{\mu\nu\rho\sigma\alpha}\tilde H_{\mu\nu}\tilde F^{(+)}_{\rho\lambda}\tilde F^{(-)}_{\sigma\lambda}\partial_\alpha\sigma\biggr].
\label{eq:finalL4}
\end{align}
Here we have defined $\tilde F^{(\pm)}\equiv e^{-2\varphi/3}F^{(\pm)}$ and $\tilde H\equiv e^{4\varphi/3}H$ along with the quartic field strength combinations $F^4\equiv F_\alpha{}^\beta F_\beta{}^\gamma F_\gamma{}^\delta F_\delta{}^\alpha$ and $ABCD\equiv A_\alpha{}^\beta B_\beta{}^\gamma C_\gamma{}^\delta D_\delta{}^\alpha$.

\subsection{Final form of the black hole solution}

After dualizing the heterotic $h$-field and Weyl scaling to the Einstein frame, and after a long chain of field redefinitions, we have finally written the four-derivative Lagrangian in a natural framework of five-dimensional supergravity coupled to two vector multiplets.  This final form of the Lagrangian, (\ref{eq:finalL2}) with (\ref{eq:finalL4}), admits a three-charge black hole solution given by
\begin{align}
    \dd s^2&=-\fft1{(H_1H_2H_3)^{2/3}}\qty(1-\alpha'\fft{7(\partial_i \log(H_1/H_2))^2-2\partial_i\log(H_1H_2)\partial_i\log H_3+19(\partial_i\log H_3)^2}{36H_3})\dd t^2\nn\\
    &\qquad+(H_1H_2H_3)^{1/3}\left(1+\alpha'\fft{(\partial_i \log(H_1/H_2))^2+\partial_i\log(H_1 H_2)\partial_i\log H_3+4(\partial_i\log H_3)^2}{18H_3}\right)\nn\\
    &\kern4em\times\left(\dd x^i\dd x^i-\alpha'\fft{\left(\partial_i\log(H_1H_2H_3)\dd x^i\right)^2 + 3\partial_{(i}\log(H_1H_2H_3)\partial_{j)}\log H_3\,\dd x^i\dd x^j}{18H_3}\right),\nn\\
    \varphi = &\;\; \dfrac{1}{4} \log(\dfrac{H_3^2}{H_1 H_2}) + \alpha'\; \dfrac{(\partial_i \log(H_1 H_2H_3))^2}{48 H_3}, \nn \\
    \sigma = & \;\; \dfrac{1}{2} \log(\dfrac{H_1}{H_2}), \nn \\
    A_t = &\;\; \dfrac{1}{H_1} \Bigg(1 + \alpha' \; \dfrac{\partial_i \log(H_1 H_2H_3)\partial_i\log(H_1H_2)}{12H_3} \Bigg), \nn \\
    B_t = &\;\;- \dfrac{1}{H_2} \Bigg(1 + \alpha' \; \dfrac{\partial_i \log(H_1 H_2H_3)\partial_i\log(H_1H_2)}{12H_3} \Bigg), \nn \\
    C_t = &\;\;\dfrac{1}{H_3} \Bigg(1 - \alpha' \; \dfrac{\partial_i\log(H_1H_2H_3)\partial_i\log H_3 }{12H_3}\Bigg).
\label{eq:finalSol}
\end{align}

At the two-derivative level, this is just the familiar three-charge STU model black hole
\begin{align}
    \dd s^2&=-\fft1{(H_1H_2H_3)^{2/3}}\dd t^2+(H_1H_2H_3)^{1/3}\dd x^i\dd x^i,\nn\\
    \varphi&=\fft14\log\left(\fft{H_3^2}{H_1H_2}\right),\qquad\sigma=\fft12\log\left(\fft{H_1}{H_2}\right),\nn\\
    A&=\fft1{H_1}\dd t,\qquad B=-\fft1{H_2}\dd t,\qquad C=\fft1{H_3}\dd t.
\end{align}
However, we see that the three charges no longer enter symmetrically at $\mathcal O(\alpha')$.  This can be understood because of the special status of the $h$-field and dilaton $\varphi$ in a string theory.  In particular, it was noted in~\cite{Liu:2023fqq} that the dilaton coupling $e^{-4\varphi/3}R_{\alpha\beta\gamma\delta}^2$ indicates that the universal vector multiplet has a special status and cannot be truncated away at this order.

The vector multiplet consisting of $(F_{\mu\nu}^{(+)},\chi^{(+)},\sigma)$ can be truncated away, as we can see from the black hole solution with $H_1=H_2$
\begin{align}
    \dd s^2&=-\fft1{(H_1^2H_3)^{2/3}}\qty(1+\alpha'\fft{4\partial_i\log H_1\,\partial_i\log H_3-19(\partial_i\log H_3)^2}{36H_3})\dd t^2\nn\\
    &\qquad+(H_1^2H_3)^{1/3}\left(1+\alpha'\fft{2\partial_i\log  H_1\,\partial_i\log H_3+4(\partial_i\log H_3)^2}{18H_3}\right)\nn\\
    &\kern4em\times\left(\dd x^i\dd x^i-\alpha'\fft{\left(\partial_i\log(H_1^2H_3)\dd x^i\right)^2 + 3\partial_{(i}\log(H_1^2H_3)\,\partial_{j)}\log H_3\,\dd x^i\dd x^j}{18H_3}\right),\nn\\
    \varphi = &\;\; \dfrac{1}{2} \log(\dfrac{H_3}{H_1}) + \alpha'\; \dfrac{(\partial_i \log(H_1^2H_3))^2}{48 H_3}, \nn \\
    \sigma = & \;\; 0, \nn \\
    A_t = -B_t=&\;\; \dfrac{1}{H_1} \Bigg(1 + \alpha' \; \dfrac{\partial_i \log(H_1^2H_3)\,\partial_i\log H_1}{6H_3} \Bigg), \nn \\
    C_t = &\;\;\dfrac{1}{H_3} \Bigg(1 - \alpha' \; \dfrac{\partial_i\log(H_1^2H_3)\,\partial_i\log H_3 }{12H_3}\Bigg).
\end{align}
However, the three-equal charge solution, which is a solution to minimal supergravity at the two-derivative level, takes the $\alpha'$-corrected form
\begin{align}
    \dd s^2&=-\fft1{H^2}\qty(1-\alpha'\fft{5(\partial_i\log H)^2}{12H})\dd t^2\nn\\
    &\qquad+H\left(1+\alpha'\fft{(\partial_i\log H)^2}{3H}\right)\left(\dd x^i\dd x^i-\alpha'\fft{\left(\partial_i\log H\,\dd x^i\right)^2}{H}\right),\nn\\
    \varphi = &\;\;  \alpha'\; \dfrac{3(\partial_i \log H)^2}{16 H}, \nn \\
    \sigma = & \;\; 0, \nn \\
    A_t = -B_t=&\;\; \dfrac{1}{H} \Bigg(1 + \alpha' \; \dfrac{\partial_i \log H\,\partial_i\log H}{2H} \Bigg), \nn \\
    C_t = &\;\;\dfrac{1}{H} \Bigg(1 - \alpha' \; \dfrac{\partial_i\log  H\,\partial_i\log H}{4H}\Bigg),
\end{align}
where $H_1=H_2=H_3=H$.  Since $\varphi$ is non-vanishing (and since the truncation $A=-B=C$ no longer holds), this is no longer a solution to minimal supergravity.  As a result, this black hole solution provides an explicit example of how the truncation to minimal supergravity can fail at the $\mathcal O(\alpha')$ level.

%%%%%%%%%%%%%%%%%%%%%%%%%%%%%%%%%%%%%%
\subsection{Properties of the three-charge solution}
\label{sec:properties}

Given the $\alpha'$ corrected three-charge black hole solution, (\ref{eq:finalSol}), it is instructive to examine some of its physical properties, and especially the effect of the $\alpha'$ corrections on its thermodynamics.  For this, we focus on a single-center solution in spherical coordinates for which the metric takes the form,
\begin{align}
    \dd s^2&=-\fft1{\mathcal H^{2/3}}\qty(1-\alpha'\fft{7(\partial_r \log(H_1/H_2))^2-2\partial_r\log(H_1H_2)\partial_r\log H_3+19(\partial_r\log H_3)^2}{36H_3})\dd t^2\nn\\
    &\qquad +\mathcal H^{1/3} \qty(1 - 2\alpha'\dfrac{\partial_r\log H_1 \partial_r\log H_2 +\partial_r\log(H_1H_2)\partial_r\log H_3}{9H_3}) dr^2\nn\\
    &\qquad+\mathcal H^{1/3}\left(1+\alpha'\fft{(\partial_r \log(H_1/H_2))^2+\partial_r\log(H_1 H_2)\partial_r\log H_3+4(\partial_r\log H_3)^2}{18H_3}\right) r^2\dd\Omega_3^2,
\label{eq:1center}
\end{align}
where the harmonic functions are defined in terms of the asymptotic charges $Q_i$,
\begin{equation}
    H_i = 1 + \dfrac{Q_i}{r^2},
\end{equation}
and where we have defined $\mathcal H=H_1H_2H_3$.

\subsubsection{Leading order}

At leading order, the solution has the well-known form
\begin{equation}
    ds^2=-\mathcal H^{-2/3}dt^2+\mathcal H^{1/3}(dr^2+r^2d\Omega_3^2).
\end{equation}
As long as all three charges are non-zero, $\mathcal H$ approaches $Q_1Q_2Q_3/r^6$ as $r\to0$.  In this case, the horizon is located at $r=0$, and the near-horizon geometry
\begin{equation}
    ds^2=-(Q_1Q_2Q_3)^{2/3}r^4dt^2+(Q_1Q_2Q_3)^{1/3}\left(\fft{dr^2}{r^2}+d\Omega_3^2\right),
\end{equation}
takes the form of $AdS_2 \times S^3$, with a horizon radius $R_H = (Q_1Q_2Q_3)^{1/6}$, as one would expect.  Note that the horizon becomes singular if any one of the charges vanishes.

Since the solution is static, one can directly read off the mass of the three-charge black hole by looking at the asymptotic behavior at infinity:
\begin{equation}
    M = \fft{\omega_3}{8\pi G_5}(Q_1+ Q_2 + Q_3),
\label{eq:bhmass}
\end{equation}
where $\omega_3=2\pi^2$ is the volume of a unit three-sphere.  Here we have restored the five-dimensional Newton's constant so that
\begin{equation}
    \mathcal L=\fft1{16\pi G_5}\sqrt{-g}(R+\cdots).
\end{equation}
That the mass is proportional to the sum of the charges is expected since the black hole is BPS.  We can also read off the leading order Bekenstein-Hawking entropy:
\begin{equation}
    S_\mathrm{BH} =\fft{A_H}{4G_5} =\dfrac{\omega_3(R_H)^3}{4G_5} = \dfrac{\pi^2}{2G_5} \sqrt{Q_1Q_2Q_3}.
\end{equation}

\subsubsection{$\mathcal O(\alpha')$ effects}

The form of the $\mathcal O(\alpha')$ corrections to the metric, (\ref{eq:1center}) is somewhat special in that they remain finite in the limit $r\to0$.  As a result, the near-horizon geometry remains $AdS_2\times S^3$, but with a shifted radius
\begin{equation}
    R_H = (Q_1Q_2Q_3)^{1/6} \qty(1 + \dfrac{2\alpha'}{3Q_3}).
\end{equation}
This area shift will shift the entropy of the black hole.  However, with the higher curvature corrections, we must also take into account the Wald entropy defined as
\begin{equation}
    S_\mathrm{W} = -2\pi\int_H \varepsilon_{\mu\nu}\epsilon_{\rho\sigma} \dfrac{\partial \mathcal{L}} {\partial R_{\mu\nu\rho\sigma}}\,\dd \Omega_3,
\label{eq:Wald}
\end{equation}
where $\varepsilon$ is the binormal to the horizon $H$. This is an antisymmetric tensor with only two nonzero components ($\varepsilon_{tr},\varepsilon_{rt}$) normalised such that $\varepsilon_{\mu\nu}\varepsilon^{\mu\nu} = -2$.  For the $\alpha'$ corrected Lagrangian given in (\ref{eq:finalL2}), we find
\begin{align}
    \dfrac{\partial \mathcal{L}} {\partial R_{\mu\nu\rho\sigma}} =& \quad \fft12(g^{\mu\rho} g^{\nu\sigma} - g^{\nu\rho} g^{\mu\sigma} )\\
    &+ \dfrac{\alpha'}{8} \left(\fft14\epsilon_{\mu\nu\alpha\beta\gamma} R_{\rho\sigma\alpha\beta} C_\gamma + 2 e^{-4\phi/3}R_{\mu\nu\rho\sigma} + \fft12 e^{-4\phi/3} (\tilde H_{\mu\nu} \tilde H_{\rho\sigma} -\tilde F^{(-)}_{\mu\nu} \tilde F^{(-)}_{\rho\sigma} )\right).
\end{align}
Substituting this into (\ref{eq:Wald}), and restoring a factor of $16\pi G_5$, gives
\begin{equation}
    S_\mathrm{W} = \dfrac{\pi^2}{2G_5} \sqrt{Q_1Q_2Q_3} \left(1 + \dfrac{2\alpha'}{Q_3} \right)\left(1- \dfrac{\alpha'}{2 Q_3} \right) = \dfrac{\pi^2}{2G_5} \sqrt{Q_1Q_2Q_3} \left(1 + \dfrac{3\alpha'}{2Q_3} \right).
\end{equation}
The $1+2\alpha'/Q_3$ correction comes from the horizon shift, while the $1-\alpha'/2Q_3$ term comes from the higher-derivative Wald entropy terms.  Since we are working to first order in $\alpha'$, the entropy can equivalently be written as
\begin{equation}
    S_\mathrm{W}=\dfrac{\pi^2}{2G_N}\sqrt{Q_1Q_2 (Q_3 + 3\alpha')}.
\end{equation}
We thus see that the $\alpha'$ correction can be viewed as a shift in the charge $Q_3$.  This agrees with the results previously obtained in \cite{Castro:2007hc,Castro:2008ys,DominisPrester:2008ynb,Elgood:2020xwu,Cano:2021nzo,Faedo:2019xii}.  While the $\mathcal O(\alpha')$ correction shifts the entropy, the mass of the black hole is unchanged from the leading order expression, (\ref{eq:bhmass}).  This is because the asymptotic mass and charges of the solution are unchanged, but of course can also be viewed as preservation of the BPS condition.

While the mass and entropy expressions above are written in terms of supergravity charges, since we are considering heterotic black holes, it is instructive to convert the asymptotic charges $Q_i$ into corresponding momentum, string and five-brane charges.  Following the approach of \cite{Cano:2021nzo}, we consider the near-horizon charges, $\tilde Q_i$, which can be read off from the $r\to0$ limit of the vector potentials.  In our case, we find
\begin{align}
    \tilde Q_1 &= Q_1 \qty(1 - \dfrac{2\alpha'}{Q_3})\nn\\
    \tilde Q_2 &= Q_2 \qty(1 - \dfrac{2\alpha'}{Q_3})\nn\\
    \tilde Q_3 &= Q_3 \qty(1 + \dfrac{\alpha'}{Q_3}).
\end{align}
From a $10$ dimensional heterotic point of view, one could think of these near-horizon charges as arising from the compactified stringy sources, namely the momentum and winding of the fundamental string and the NS$5$-brane charge. Following the conventions of \cite{Cano:2021nzo} we parametrise these respectively by the integers $n,w,N$. However, note that we performed a total derivative shift to the torsionful Cherns-Simons form in $10$ dimensions \cite{Jayaprakash:2024xlr}, which causes a shift to the five-dimensional gauge field $B$
\begin{equation}
    \delta B_\mu = \dfrac{\alpha'}{4} e^\sigma\left(\dfrac{1}{2} \omega_{+\mu}^{\alpha\beta} F^{(-)\alpha\beta}+ F_{\mu\alpha}^{+}\partial_\alpha\sigma\right).
\end{equation}
In the near horizon limit, this is reflected as a shift on the charge $\tilde Q_2$,
\begin{equation}
    \delta \tilde Q_2 =  -2\alpha'\dfrac{Q_2}{Q_3}.
\end{equation}
So we would expect $\hat Q_2 = \tilde Q_2 - \delta\tilde Q_2$ to have the correct string quantisation. So in the conventions of \cite{Cano:2021nzo}, we have
\begin{equation}
    \tilde Q_1 = \dfrac{g_s^2 \alpha'{}^2}{R_z^2}n, \qquad \hat{Q}_2 = g_s^2 \alpha' w , \qquad \tilde Q_3 = \alpha'N, \qquad G_5 = \dfrac{\pi g_s^2 \alpha'{}^2}{4R_z},
\end{equation}
where $R_z$ is the radius of the compactification circle.  As a result, the entropy can be given in terms of the quantized string charges as
\begin{equation}
    S_W = 2\pi \sqrt{nw(N+4)}.
\end{equation}
%

%%%%%%%%%%%%%%%%%%%%%%%%%%%%%%%%%%%%%%
\section{Discussion}\label{sec:disc}

In this paper, we have explored the structure of multicenter BPS black hole solutions to heterotic supergravity.  In six dimensions and above, the basic solution is a two-charge black hole~\cite{Ruiperez:2020qda,Cano:2021dyy}, which takes the relatively simple form, (\ref{eq:twocharge}), when written in a standard Kaluza-Klein basis.  Of course, one of the issues facing any study of higher-derivative corrections is that of field redefinitions.  Since we take a perturbative approach, one may always choose alternate field redefinition bases according to preferences and for highlighting various aspects of the theory.  In this sense, the same two-charge black hole solution can be written as (\ref{eq:fr2cbh}) when working in the basis where all derivatives of field strengths such as $\nabla F$ have been removed.

While the field-redefined Lagrangian, (\ref{eq:4derlag3}), is perhaps more natural from the point of view of the lower-dimensional supergravity, the corresponding black hole solution, (\ref{eq:fr2cbh}), is now more complicated, as the $\mathcal O(\alpha')$ terms are no longer captured by a single quantity $T$ given in (\ref{eq:Tcombo}).  Moreover, the multicenter solution is no longer written in strictly isotropic coordinates, and in this case, the spatial metric takes the form
\begin{equation}
    g_{ij}=\delta_{ij}-\fft{\alpha'}8\partial_i\log(H_1H_2)\,\partial_j\log(H_1H_2).
\label{eq:noniso}
\end{equation}
In general, the construction of multicenter BPS black holes starts with an isotropic coordinate ansatz of the form
\begin{equation}
    \dd s^2=-e^{2f}\dd t^2+e^{2g}\dd x^i\dd x^i.
\end{equation}
The resulting solution is then written in terms of a set of harmonic functions on the flat transverse Euclidean space.  While this generally holds at leading order, the field-redefined solution, (\ref{eq:noniso}), demonstrates that the isotropic coordinate ansatz may no longer be appropriate once higher-derivative corrections are included.  As we have seen, whether isotropic coordinates can be used or not depends on the particular choice of field redefinition frame, as the original solution, (\ref{eq:twocharge}), does have an isotropic form.  Nevertheless, bottom-up constructions of higher-derivative BPS black holes may benefit from the use of a non-isotropic coordinate ansatz.

As BPS black holes are protected by supersymmetry, it is perhaps not surprising that many of their features persist when including higher-derivative corrections.  In particular, the supersymmetry algebra guarantees the existence of shortened representations with $M=Q$ regardless of its realization in a higher-derivative theory.  This suggests that the force balance condition is still maintained, thus allowing for static multicentered BPS black hole configurations.  It is nevertheless remarkable that the $\mathcal O(\alpha')$ corrections can be written in a relatively simple manner in terms of the leading order harmonic functions.  This suggests some form of universality in the harmonic function construction of BPS configurations.  It would be interesting to see if this framework continues to hold to higher orders in the derivative expansion.

\subsection{BPS black hole in five-dimensional supergravity}

Compactification to five dimensions is particularly interesting, as the final form of the three-charge black hole solution, (\ref{eq:finalSol}), fits within the framework of five-dimensional $\mathcal N=2$ supergravity and the extensively studied STU model.  From the heterotic point of view, this model may be obtained by compactification on $K3\times S^1$ or, in our approach, on $T^4\times S^1$ followed by additional truncation.

At the two-derivative level, the bosonic STU model Lagrangian can be written in the language of very special geometry as
\begin{equation}
    e^{-1}\mathcal L_{\partial^2}=R-G_{IJ}\partial_\mu X^I\partial_\mu X^J-\fft12G_{IJ}F_{\mu\nu}^IF_{\mu\nu}^J+\fft14C_{IJK}\epsilon^{\mu\nu\rho\sigma\lambda}F_{\mu\nu}^IF_{\rho\sigma}^JA_\lambda^K,
\end{equation}
where the constrained scalars $X^I$ satisfy
\begin{equation}
    \mathcal C\equiv\fft16C_{IJK}X^IX^JX^K=1,
\end{equation}
and
\begin{equation}
    G_{IJ}=-\fft12\partial_I\partial_J\log\mathcal C.
\end{equation}
In particular, the STU model is obtained by taking $C_{123}=1$, and the two derivative part of the Lagrangian, (\ref{eq:finalL2}), corresponds to setting
\begin{align}
    &X^1=e^{\fft23\varphi-\sigma}, &&X^2=e^{\fft23\varphi+\sigma}, &&X^3=e^{-\fft43\varphi},\nn\\
    &F_{\mu\nu}^1=F_{\mu\nu}, &&F_{\mu\nu}^2=-G_{\mu\nu}, &&F_{\mu\nu}^3=H_{\mu\nu}.
\end{align}

Curvature-squared corrections in $\mathcal N=2$ supergravity have been extensively studied and constructed using superspace and superconformal tensor calculus methods~\cite{Hanaki:2006pj,Bergshoeff:2011xn,Ozkan:2013uk,Ozkan:2013nwa,Liu:2022sew,Gold:2023dfe,Gold:2023ymc,Gold:2023ykx,Cassani:2024tvk} (see~\cite{Ozkan:2024euj} for a recent review).  While minimal $\mathcal N=2$ supergravity has a unique on-shell four-derivative invariant, multiple off-shell invariants for supergravity coupled to vector multiplets have been constructed involving Riemann-squared and Ricci-squared couplings.  The standard Weyl-squared invariant of~\cite{Hanaki:2006pj} is perhaps the most familiar, and its auxiliary fields were recently integrated out in~\cite{Cassani:2024tvk}.  The result is that, in addition to the cubic prepotential given through $C_{IJK}$, the $R^2$ corrections are parametrized by a set of couplings $\lambda_I$ of the form~\cite{Hanaki:2006pj,Cassani:2024tvk}
\begin{equation}
    e^{-1}\mathcal L_{\partial^4}=\lambda_IX^IR_{\mu\nu\rho\sigma}^2+\fft12\lambda_I\epsilon^{\mu\nu\rho\sigma\lambda}R_{\mu\nu\alpha\beta}R_{\rho\sigma}{}^{\alpha\beta}A_\lambda^I+\cdots.
\label{eq:cassani}
\end{equation}
Comparison with (\ref{eq:finalL4}) indicates that the compactified heterotic Lagrangian corresponds to taking
\begin{equation}
    \lambda_1=0,\qquad\lambda_2=0,\qquad\lambda_3=\fft{\alpha'}8.
\label{eq:lambdas}
\end{equation}
Note that $\lambda_3$ corresponds to the $h$-field coupling from the heterotic point of view.  As we have seen above, the dilaton and $h$-field are part of a universal vector multiplet that cannot be truncated out at the four-derivative level.

As the reduced four-derivative heterotic action, (\ref{eq:finalL4}), fits the structure of the STU model with the identification (\ref{eq:lambdas}), one may naturally expect that it would agree with the corresponding bottom-up supergravity Lagrangian, (\ref{eq:cassani}).  However, here we end up with a surprise in that the heterotic and supergravity Lagrangians disagree in terms that do not involve the Riemann tensor.  Of course, whenever we are comparing higher-derivative Lagrangians, we must take into account the possibility of field redefinitions.  However, in the present case, note that the heterotic action, (\ref{eq:finalL2}) and (\ref{eq:finalL4}) is by construction even in $H$ in the CP-even sector and odd in $H$ in the CP-odd sector. Moreover, field redefinitions using the two-derivative equations of motion cannot change this structure.  On the other hand, inserting the STU parameters into the four-derivative action, (\ref{eq:cassani}), of~\cite{Cassani:2024tvk} yields both even and odd $H$ terms in all sectors of the Lagrangian.  This demonstrates that the reduced heterotic Lagrangian is not contained in the four-derivative superinvariant of~\cite{Hanaki:2006pj,Cassani:2024tvk}.

The question then naturally arises as to how to interpret the heterotic four-derivative coupling from an $\mathcal N=2$ supergravity point of view.  One possibility is that in fact two Riemann-squared invariants have been constructed --- one using the standard Weyl multiplet~\cite{Hanaki:2006pj} and one using the dilaton-Weyl multiplet~\cite{Bergshoeff:2011xn}.  In this case, the heterotic invariant may be a combination of these two invariants.  Another possibility is that whenever we are working with supergravity coupled to vector multiplets, there can also be four-derivative vector multiplet invariants of the form $F^4$ and its completion.  In fact, by suitable field redefinitions, the Ricci-squared invariants essentially reduce to this case.

There is of course the exciting possibility that the space of four-derivative $\mathcal N=2$ invariants has yet to be exhausted and that the heterotic theory corresponds to a new five-dimensional supergravity invariant.  Since these higher-derivative corrections are at the core of precision holography even beyond the BPS sector, it is of essential importance to complete the picture of five-dimensional superinvariants.  This is currently under investigation and will be reported on in an upcoming work~\cite{Cai:2024}.

\section*{Acknowledgements}
This work was supported in part by the U.S. Department of Energy under grant DE-SC0007859. SJ was supported in part by a Leinweber Graduate Summer Fellowship. RJS was supported in part by a Leinweber Summer Research Award and by the National Key Research and Development Program No. 2022YFE0134300.

%%%%%%%%%%%%%%%%%%%%%%%%%%%%%%%%%%%%%%%%%%%%%%%%%%%%%%%%%%%%%%%%%%%%%%

\bibliographystyle{JHEP}
\bibliography{cite}

\end{document}